\documentclass[preprint2]{proto}
\usepackage{times}
\newcommand{\refs}{\par\noindent\hangindent=1pc\hangafter=1}

\begin{document}

\def\kms{{km\thinspace s$^{-1}$}}
\def\13co{$^{13}$CO}
\def\h2{H$_2$}
\def\l{$l$}
\def\le{{$l$}\ }
\def\b{$b$}
\def\be{{$b$}\ }
\def\lesssim{\mathrel{\hbox{\rlap{\hbox{\lower4pt\hbox{$\sim$}}}\hbox{$<$}}}}
\def\gtrsim{\mathrel{\hbox{\rlap{\hbox{\lower4pt\hbox{$\sim$}}}\hbox{$>$}}}}
\let\la=\lesssim
\let\ga=\gtrsim
\def\deg{\ifmmode^\circ\else$^\circ$\fi}
\def\dege{{\ifmmode^\circ\else$^\circ$\fi}\ }
\def\solar{\ifmmode _{\mathord\odot}\else$_{\mathord\odot}$\fi}
\def\sune{{$_\odot$}\ }
\def\rsun{R\sun}
\def\tsun{$\Theta$\sun}
\def\msun{M\sun}
\def\rsune{{R\sun}\ }
\def\msune{{M\sun}\ }
\def\tsune{{$Thets$\sun}\ }
\def\asece{$^{\prime \prime}$\ }
\def\asec{$^{\prime \prime}$}
\def\amine{$^{\prime}$\ }
\def\amin{$^{\prime}$}
\def\ete{{\it et al.}\ } %et al.%
\def\et{{\it et al.}}
\def\etal{{\it et al.}\ } %et al.%
\def\ege{{\it e.g.}\ }   %e.g.%
\def\eg{{\it e.g.}}
\def\iee{{\it i.e.}\ }   %i.e.%
\def\ie{{\it i.e.}}
\def\microne{{$\mu$m}\ }
\def\micron{{$\mu$m}}
\def\hper{\ifmmode \rlap.^{h} \else $\rlap{.}^h$\fi}
                                %Superscript 'h' over period%
\def\mper{\ifmmode \rlap.^{m} \else $\rlap{.}^m$\fi}
                                %Superscript 'm' over period%
\def\sper{\ifmmode \rlap.^{s} \else $\rlap{.}^s$ \fi}
                                %Superscript 's' over period.%
\def\degper{\ifmmode \rlap.^{\circ} \else $\rlap{.}^{\circ} $\fi}
                                                %' over period.%
\def\arcmper{\ifmmode \rlap.{' } \else $\rlap{.}' $\fi}
                                                %' over period.%
\def\arcsper{\ifmmode \rlap.{'' } \else $\rlap{.}'' $\fi}
                                                %'' over period.%
\def\cc{cm$^{-3}$}
\def\cce{cm$^{-3}$\ }
\def\c2{cm$^{-2}$}
\def\2e{$^{2}$\ }
\def\2{$^{2}$}
\def\3e{$^{3}$\ }
\def\3{$^{3}$}
\def\-1e{$^{-1}$\ }
\def\-1{$^{-1}$}
\def\-2e{$^{-2}$\ }
\def\-2{$^{-2}$}
\def\-3e{$^{-3}$\ }
\def\-3{$^{-3}$}
\def\grapprox{$_>\atop{^\sim}$}
                %Greater than over approximately.%
\def\lapprox{$_<\atop{^\sim}$}  %Less than over approximately.%
\newcommand{\co}{\mbox{\rm CO}}
\newcommand{\cothree}{{\rm $^{13}$CO}}
\newcommand{\cn}{{\rm C$^0$}}
\newcommand{\cp}{{\rm C$^+$}}
\newcommand{\ci}{\mbox{\rm [\ion{C}{1}]}}
\newcommand{\oi}{\mbox{\rm [\ion{O}{1}]}}
\newcommand{\cii}{\mbox{\rm [\ion{C}{2}]}}
\newcommand{\oiii}{\mbox{\rm [\ion{O}{3}]}}
\newcommand{\hi}{\mbox{\rm \ion{H}{1}}}
\newcommand{\hii}{\mbox{\rm \ion{H}{2}}}
\newcommand{\htwo}{\mbox{\rm H$_2$}}
\newcommand{\jone}{\mbox{($1\rightarrow0$)}}
\newcommand{\jtwo}{\mbox{($2\rightarrow1$)}}
\newcommand{\jthree}{($3\rightarrow2$)}
\newcommand{\jfour}{($4\rightarrow3$)}
\newcommand{\fsone}{($^3$P$_1\rightarrow^3$P$_0$)}
\newcommand{\fstwo}{($^3$P$_2\rightarrow^3$P$_1$)}
\newcommand{\fseno}{($^3$P$_0\rightarrow^3$P$_1$)}
\newcommand{\fsowt}{($^3$P$_1\rightarrow^3$P$_2$)}
\newcommand{\fscii}{($^2$P$_{3/2}\rightarrow^2$P$_{1/2}$)}
\newcommand{\percmcu}{cm$^{-3}$}
\newcommand{\percmsq}{cm$^{-2}$}
\newcommand{\persr}{sr$^{-1}$}
\newcommand{\kmpers}{\mbox{km~s$^{-1}$}}
\newcommand{\Kkmpers}{\mbox{K~km~s$^{-1}$}}
\newcommand{\intunits}{erg~s$^{-1}$~cm$^{-2}$~sr$^{-1}$}
\newcommand{\intensity}{erg~s$^{-1}$~cm$^{-2}$}
\newcommand{\fluxdens}{erg~s$^{-1}$~cm$^{-2}$~$\mu$m$^{-1}$}
\newcommand{\xcounits}{\mbox{cm$^{-2}$ (K km s$^{-1}$)$^{-1}$}}
\newcommand{\Ico}{\mbox{\rm I$_{\rm CO}$}}
\newcommand{\Ici}{\mbox{\rm I$_{\rm [CI]}$}}
\newcommand{\Icii}{\mbox{\rm I$_{\rm [CII]}$}}
\newcommand{\Ifir}{\mbox{\rm I$_{\rm FIR}$}}
\newcommand{\citoco}{\Ici/\Ico}
\newcommand{\ciitoco}{\Icii/\Ico}
\newcommand{\ciitoci}{\Icii/\Ici}
\newcommand{\xco}{\ensuremath{X_{\mathrm{CO}}}}
\newcommand{\aco}{\mbox{$\alpha_{\rm CO}$}}
\newcommand{\av}{\mbox{\rm A$_{\rm V}$}}
\newcommand{\xuv}{\mbox{$\chi_{\rm uv}$}}
\newcommand{\vlsr}{\mbox{V$_{\rm LSR}$}}
\renewcommand{\pasp}{{\it Publ. Astron. Soc. Pac.}}%
\renewcommand{\nat}{{\em Nature}}
\title{\textbf{\LARGE Giant Molecular Clouds in Local Group Galaxies}}

%\author {\textbf{\large Leo Blitz, Yasuo Fukui, Akiko Kawamura, Adam Leroy,
%Norikazu Mizuno, Erik Rosolowsky}}
%\affil{\small\em University of California, Berkeley, University of Nagoya, 
%Harvard-Smithsonian Center for Astrophysics}
\author {\textbf{\large Leo Blitz}}
\affil{\small\em University of California, Berkeley}
\author {\textbf{\large  Yasuo Fukui}} 
\affil{\small\em Nagoya University}
\author {\textbf{\large  Akiko Kawamura}}
\affil{\small\em  Nagoya University}
\author {\textbf{\large  Adam Leroy}}
\affil{\small\em University of California, Berkeley}
\author {\textbf{\large  Norikazu Mizuno}}
\affil{\small\em  Nagoya University}
\author {\textbf{\large  Erik Rosolowsky}}
\affil{\small\em  Harvard-Smithsonian Center for Astrophysics}

\begin{abstract} 
\baselineskip = 11pt 
\leftskip = 0.65in 
\rightskip = 0.65in \parindent=1pc {\small We present the first
comparative study of extragalactic GMCs using complete data sets for
entire galaxies and a uniform set of reduction and analysis
techniques.  We present results based on CO observations for the LMC,
SMC, M33, M31, IC10 and the nucleus of M64, and make comparisons with
archival Milky Way observations.  Our sample includes large spirals
and dwarf irregulars with metallicities that vary by an order of
magnitude. GMCs in \ion{H}{1} rich galaxies are seen to be
well-correlated with \ion{H}{1} filaments that pervade the galactic
disks, suggesting that they form from pre-existing \ion{H}{1}
structures. Virial estimates of the ratio of CO line strength to \h2
column density, \xco, suggests that a value of 4 $\times 10^{20}$
cm$^{-2} (\Kkmpers)^{-1}$ is a good value to use in most galaxies
(except the SMC) if the GMCs are virialized.  However, if the clouds
are only marginally self-gravitating, as appears to be the case
judging from their appearance, half the virial value may be more
appropriate. There is no clear trend of $\xco$ with metallicity.  The
clouds within a galaxy are shown to have the about the same \h2
surface density and differences between galaxies seem to be no more
than a factor of $\sim 2$.  We show that hydrostatic pressure appears
to be the main factor in determining what fraction of atomic gas is
turned into molecules.  In the high-pressure regions often found in
galactic centers, the observed properties of GMCs appear to be
different from those in the found in the Local Group. From the
association of tracers of star formation with GMCs in the LMC, we find
that about 1/4 of the GMCs exhibit no evidence of star formation and
we estimate that the lifetime of a typical GMC in these galaxies is
20--30 Myr.
%This is probably an upper limit because of our
%insensitivity to low mass star forming systems such as 
%the Taurus and Ophiuchus.  
\\~\\~\\~} 
\end{abstract}

\section{\textbf{INTRODUCTION}}

Although a great deal of progress has been made on the topic of star
and planet formation since the last Protostars and Planets conference
in Santa Barbara, little work has been done to connect what we know
about star formation in the Milky Way to star formation in the
Universe as a whole.  Fundamental limitations include only a weak
understanding of how the massive stars form, how clusters and
associations form, and the constancy of the IMF.  After all, in
external galaxies, we generally observe only the effects of massive
star formation and the formation of star clusters.  Furthermore,
knowledge of the initial conditions for star formation at all masses
remains elusive both within and outside of the Milky Way.

\begin{deluxetable}{l l c c c}
\tabletypesize{\small}
\tablewidth{0pt}
\tablecolumns{5}
\tablecaption{\label{DATATAB} Local Group GMC Data}
\tablehead{ \colhead{Galaxy} & \colhead{Telescope} &
\colhead{Metallicity} & \colhead{Spatial Resolution} & \colhead{Reference}}
\startdata
LMC & NANTEN & $0.33~Z_\odot$ & $40$ pc & 1 \\
SMC & NANTEN & $0.1~Z_\odot$ & $48$ pc & 2 \\
IC10 & OVRO/BIMA & $0.25~Z_\odot$ & $14$ -- $20$ pc & 3\\
M33 & BIMA & $0.1$ -- $1.0~Z_\odot$ & $20$ -- $30$ pc & 4\\
M31 & BIMA & $0.5~Z_\odot$ & $26$ -- $36$ pc & 5\\
\enddata
\tablerefs{(1) {\it Fukui et al.}~(2006) (2) {\it Mizuno et al.}~(2006) 
(3) {\it Leroy et al.}~(2006) (4) {\it Engargiola et al.}~(2003) 
(5) {\it Rosolowsky} (2006)}
\end{deluxetable}

Since nearly all stars form in Giant Molecular Clouds (GMCs), one way
to make progress is to examine the properties of GMCs in a number of
different extragalactic environments to see how they differ.  From the
similarities and differences, it might be possible to make some
general conclusions about how star formation varies throughout the
Universe.  Although individual, extragalactic GMCs had been observed
previously at high enough resolution to at least marginally resolve
them (e.g.,~ {\it Vogel et al.}, 1987; {\it Lada et al.}, 1988), the
first attempts to do this in a systematic way were by Christine Wilson
({\it Wilson and Scoville}, 1990; {\it Wilson and Reid}, 1991; {\it
Wilson and Rudolph}, 1993; {\it Wilson}, 1994) using the OVRO and BIMA
interferometers. Her efforts were hampered by small survey areas in a
few galaxies, so general conclusions could only be made by
extrapolation.  Numerous other authors subsequently studied one or a
few extragalactic GMCs, both in the Local Group and beyond.  An
exhaustive list of their efforts is beyond the scope of the present
article.

The situation has changed in the last five years as a result of the
construction of the NANTEN telescope in the Southern Hemisphere and
the completion of the 10-element BIMA Array.  The former made it
possible to map the Magellanic Clouds completely with high enough
spatial resolution and signal-to-noise to identify all of the GMCs
with masses ~$ > 3 \times 10^4$ $M_\sun$; the completion of the BIMA
interferometer made it possible to identify GMCs in other, more
distant galaxies in the Local Group. Because of their relatively large
fields of view, these two telescopes could completely survey nearby
galaxies. Thus, the first complete survey of GMCs in {\it any} galaxy
was of the LMC ({\it Fukui et al.},~1999; {\it Mizuno et al.},~2001b)
and not the Milky Way (MW).  Although the molecular gas in the MW has
been essentially completely mapped, velocity crowding in many
directions makes it impossible to generate a full catalog of GMCs.
Similarly, the first complete CO surveys of the Magellanic Clouds were
by {\it Cohen et al.}~(1988) and {\it Rubio et al.}~(1991), but the
resolution was too poor to determine the properties of individual
molecular clouds.

In this paper, we review the recent surveys of CO in Local Group
galaxies that (1) have sufficient resolution to study individual
molecular clouds and (2) span all or most of the target galaxy.  We
compare the results of observations of GMCs in the four external Local
Group galaxies that have been mapped in their entirety in CO: the
Large Magellanic Cloud (LMC, {\it Fukui et al.},~2001; {\it Fukui et
al.},~2006), the Small Magellanic Cloud (SMC, {\it Miznuo et al.},
2001a; {\it Mizuno et al.},~2006), IC~10 ({\it Leroy et al.},~2006),
and M33 ({\it Engargiola et al.},~2003). We have also made
observations in a small strip in M31 ({\it Rosolowsky}, 2006), and we
compare the properties of the GMCs in all of these galaxies to clouds
in the outer MW (from {\it Dame et al.},~2001) using a uniform set of
analytic techniques.  The LMC and SMC observations were made with the
single-dish NANTEN telescope in Chile, the remaining galaxies were
observed with the BIMA millimeter-wave interferometer at Hat Creek,
California (combined with obsevations from the Caltech OVRO millimeter
interferometer for IC~10).  A tabulation of the galaxies we observed,
their metallicities and the resolution used to observe them is given
in Table 1.

% In
%addition, we make comparisons of the GMCs with those in the nucleus of
%the galaxy M64.  In that galaxy, the central surface density over the
%inner 500 pc radius is comparable to the surface density of a typical
%Local Group GMC. A summary of the galaxies we observed is given in
%Table 1.

\bigskip
\noindent
\section {\textbf{THE GALAXIES }}
\bigskip

In this section, we examine the distribution of CO emission in the
surveyed galaxies and we compare the CO to emission in other
wavebands.

\subsection{The LMC}
\begin{figure}[h!]
\epsscale{1.0}
\plotone{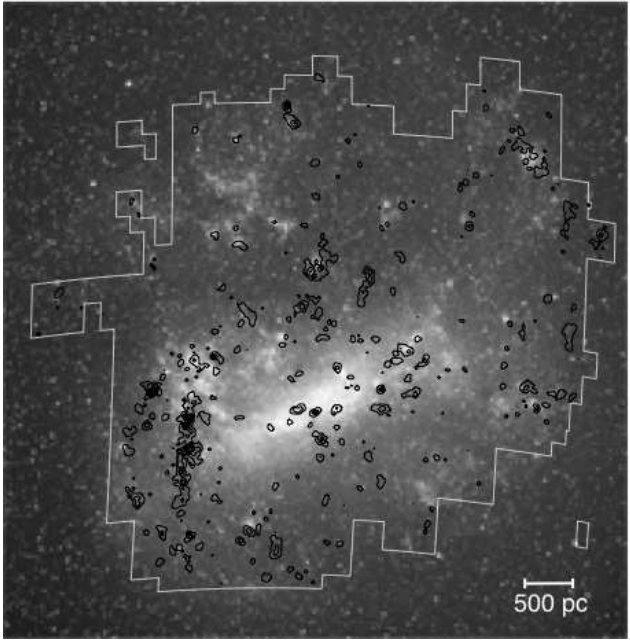} \caption{\small Optical image of the LMC
 with GMCs mapped with the NANTEN telescope indicated within the
 boundary of the survey area.  The CO is well correlated with
 \ion{H}{2} regions.  The GMCs are easily identified by eye except for
 the region south of 30 Doradus where they appear as a vertical line
 of clouds and the individual GMCs may be overlapping
 in this region.\label{lmcmap}}
\end{figure}

Fig.~\ref{lmcmap} shows the molecular clouds detected with the NANTEN
Survey ({\it Fukui et al.},~2001; {\it Fukui et al.},~2006) on an
optical image of the LMC.  Except for a region near the eastern edge
of the galaxy (left side of Fig.~\ref{lmcmap}) below 30 Doradus, the
clouds appear to be spatially well-separated and it is possible to
pick them out individually by eye.  The long string of bright CO
emission along the eastern edge of the galaxy is likely composed of
several clouds that cannot be separated at this resolution.  Some have
speculated that this feature is due to hydrodynamical collision
between the LMC and SMC ({\it Fujimoto and Noguchi}, 1990) or ram
pressure pileup of gas due to the motion of the LMC through a halo of
hot, diffuse gas ({\it de Boer et al.},~1998; {\it Kim et al.},~1998).
Supershells may also be playing a role in the formation of GMCs as in
the case LMC4 ({\it Yamaguchi et al.},~2001a).  A comprehensive
comparison between supergiant shells and GMCs shows that only about
1/3 of the GMCs are located towards supershells, suggesting the
effects of supershells are not predominant ({\it Yamaguchi et
al.},~2001b). There is neither an excess nor a deficit of CO
associated with the stellar bar, but the bright \ion{H}{2} regions are
all clearly associated with molecular clouds.  Individual clouds are
frequently associated with young clusters of stars.  Not every cluster
of young stars is associated with a cloud nor does every cloud show
evidence of massive star formation.  Using this association and the
ages of the stellar clusters, we can establish the evolutionary time
scale for GMCs (Section 
\ref{timescale}).

\subsection{The SMC}

\begin{figure}[hb!]
\epsscale{1.0}
 \plotone{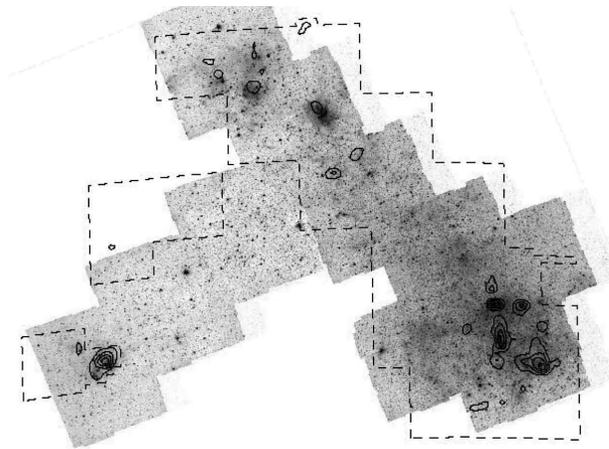} \caption{\small GMCs in the SMC observed
 with the NANTEN telescope overlayed on an a near-infrared image of
 the galaxy from the {\it Spitzer} Space Telescope ({\it Bolatto et
 al.},~2006).  The lines indicate the survey boundary.  The CO clouds
 are clearly associated with regions of transiently heated small
 grains or PAHs that appear as dark, nebulous regions in the
 image. \label{smcmap}}
\end{figure}

Fig.~\ref{smcmap} shows the GMCs superimposed on a grayscale image
made using the 3.6, 4.5, and 8.0 $\mu$m bands from the IRAC instrument
on the {\it Spitzer} Space Telescope ({\it Bolatto et al.},~2006). The
CO map is from the NANTEN telescope ({\it Mizuno et al.},~2006).  As in
the LMC, the GMCs in the SMC are easily identified by eye. Unlike the
LMC, they are not spread throughout the galaxy but appear
preferentially on the northern and southern ends of the galaxy.
Another grouping is located to the east (left) of the SMC along the
\ion{H}{1} bridge that connects the LMC and SMC, apparently outside
the stellar confines of the galaxy.  The {\it Spitzer} image traces
the stellar continuum as well as warm dust and PAH emission.  The 8.0
$\mu$m emission is associated with the molecular gas traced by CO, but
appears to be more extended than the CO emission.  The SMC has the
lowest metallicity in our sample and provides an opportunity to
explore the behavior of molecular gas in chemically primitive
environments.

%The right hand side of Figure 2 shows the molecular clouds overlayed
%on an \ion{H}{1} image of the SMC.  As in the LMC, the GMCs are
%associated with local maxima in the \ion{H}{1} emission.  However, the
%\ion{H}{1} emission seems somewhat more diffuse and less filamentary
%than in the LMC.

\subsection{M33}
\begin{figure}[t!]
\epsscale{1.0}
 \plotone{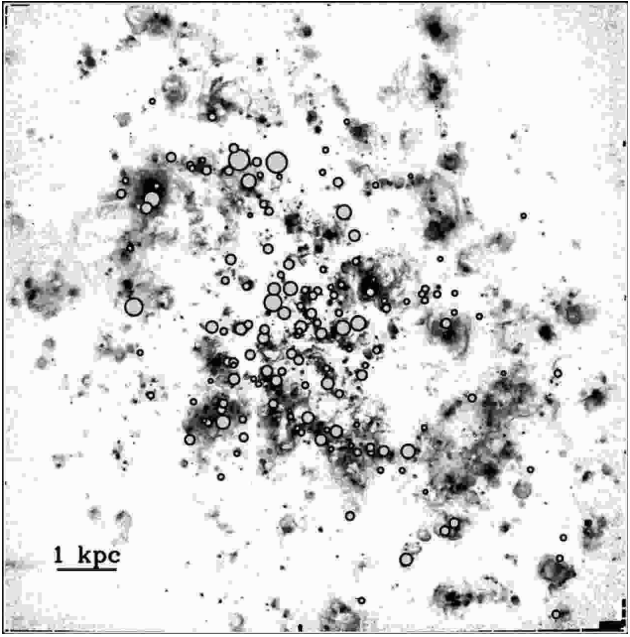} \caption{\small The locations of GMCs in M33
 as derived from the 759 field BIMA mosaic of {\it Engargiola et
 al.}~(2003).  Since sources of CO emission in a map would be too
 small to identify in the figure, the locations of GMCs are instead
 indicated by light gray circles.  The area of the circles is scaled to the
 CO luminosity which should be proportional to the H$_2$ mass.  The
 GMC locations are overlayed on a continuum subtracted H$\alpha$ image
 of the galaxy ({\it Massey et al.}, 2001).  There is significant
 correlation between the GMCs and massive star formation as traced by
 H$\alpha$.
\label{m33map}}
\end{figure}

Fig.~\ref{m33map} shows the locations of GMCs in M33 from the BIMA
telescope ({\it Engargiola et al.},~2003) superimposed on an H$\alpha$ image
of the galaxy ({\it Massey et al.},~2001).  The two low-contrast spiral arms
({\it Regan and Vogel}, 1994) are well-traced by GMCs, but the GMCs are not
confined to these arms as is evident in the center of the galaxy.
There is good spatial correlation between the GMCs and the \ion{H}{2}
regions.  Once again, the correlation is not perfect and there are
GMCs without \ion{H}{2} regions and vice versa.  Unlike the other
images, we show the locations of the GMCs as circles with areas
proportional to the CO luminosity of each GMC; the CO luminosity is
expected to be proportional to the H$_2$ mass of each GMC.  Note that
the most massive GMCs ($\sim 10^6~M_{\odot}$) are not found toward the
center of the galaxy but along spiral arms north of the galactic
nucleus.  These massive clouds are relatively devoid of H$\alpha$
emission. The completeness limit of this survey is about 1.5~$\times
10^5~M_\sun$; thus there are presumably many lower mass clouds below
the limit of sensitivity.  Many of these low mass clouds are likely
associated with the unaccompanied \ion{H}{2} regions in the figure.

%The right hand panel shows the GMCs superimposed on the \ion{H}{1} image of
%the galaxy made by Deul and van der Hulst (1987).  Several things are
%striking about this image.  First, the \ion{H}{1} is quite filamentary, and
%the GMCs are strongly confined to the filaments.  Second, although the
%filaments have nearly constant column density with radius, very few
%GMCs are found at galactic radii beyond 4 kpc.  Apparently, \ion{H}{1} column
%density alone is not a sufficient criterion for determining the location
%of GMCs in galaxies.  Third, the large relatively empty regions, the
%\ion{H}{1} ``holes,'' are not evacuated by explosions. The large ones require
%about 10$^{53}$ ergs to evacuate, but there are no obvious stellar
%clusters remaining at the center of the holes.  Furthermore, x-ray
%emission is not concentrated in the holes.  The large holes are thus
%probably defined by the filaments, which are likely to have a
%gravitational, or density wave origin.  Small holes with d $<$ 200pc, 
%on the other hand, are found to be well correlated with OB
%associations (Deul and van der Hulst 1987); these tend to be
%concentrated toward the center of the galaxy. 

\subsection{IC 10}
\begin{figure}[ht!]  
\epsscale{1.0} \plotone{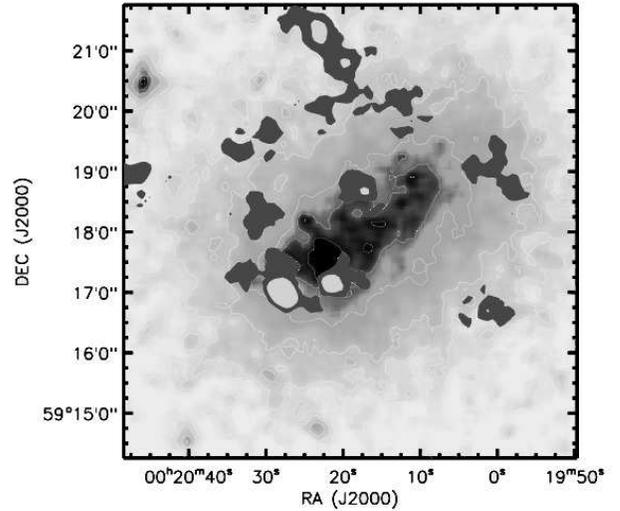} 
\caption{\small The GMCs in IC~10 made from a 50 field mosaic of the
galaxy with the BIMA telescope overlayed on a 2 $\mu$m image of the
galaxy from the 2MASS survey. The dark gray area 
corresponds to CO brightness above 1 K \kms; the light gray area
corresponds to CO brightness above 10 K \kms. The black region in the
center has the highest stellar surface density.  The rms noise of the CO
data is $\sim$ 0.3 K \kms in each channel map; the peak value in the
integrated intensity map is 48 K \kms ({\it Leroy et al.},~2006).
\label{ic10map}}  
\end{figure}

Fig.~\ref{ic10map} is an image of the GMCs in IC~10 from a 50 field CO
mosaic with the BIMA telescope ({\it Leroy et al.},~2006) superimposed on a 2
$\mu$m image of the galaxy made from 2MASS data ({\it Jarrett et al.},~2003).
As with the Magellanic Clouds and M33, the GMCs show no obvious
spatial correlation with old stellar population -- some massive clouds
are found where there are relatively few stars.

\subsection{The Correlation with \ion{H}{1}}

\begin{figure*}
\epsscale{2.25}
\plottwo{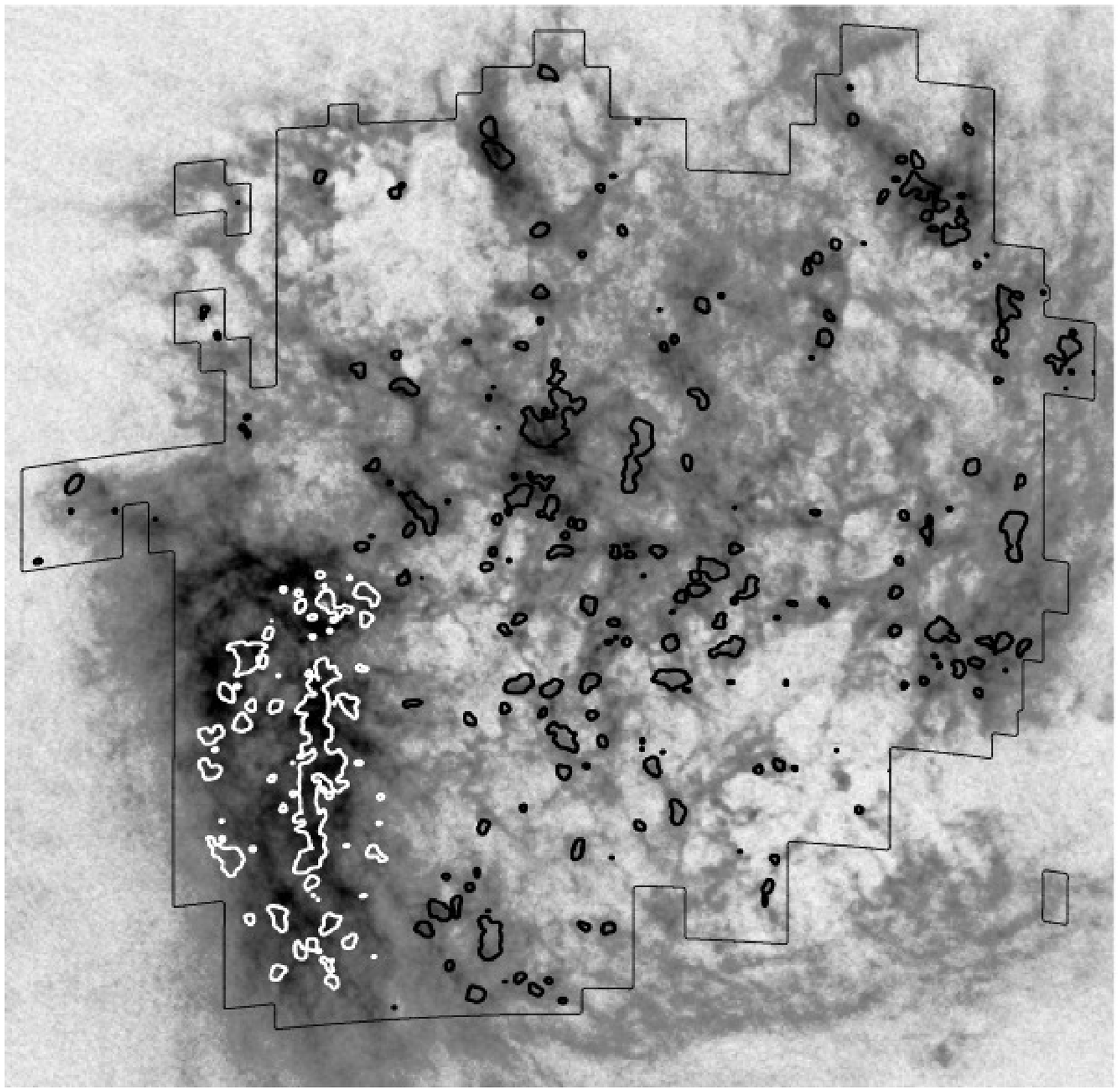}{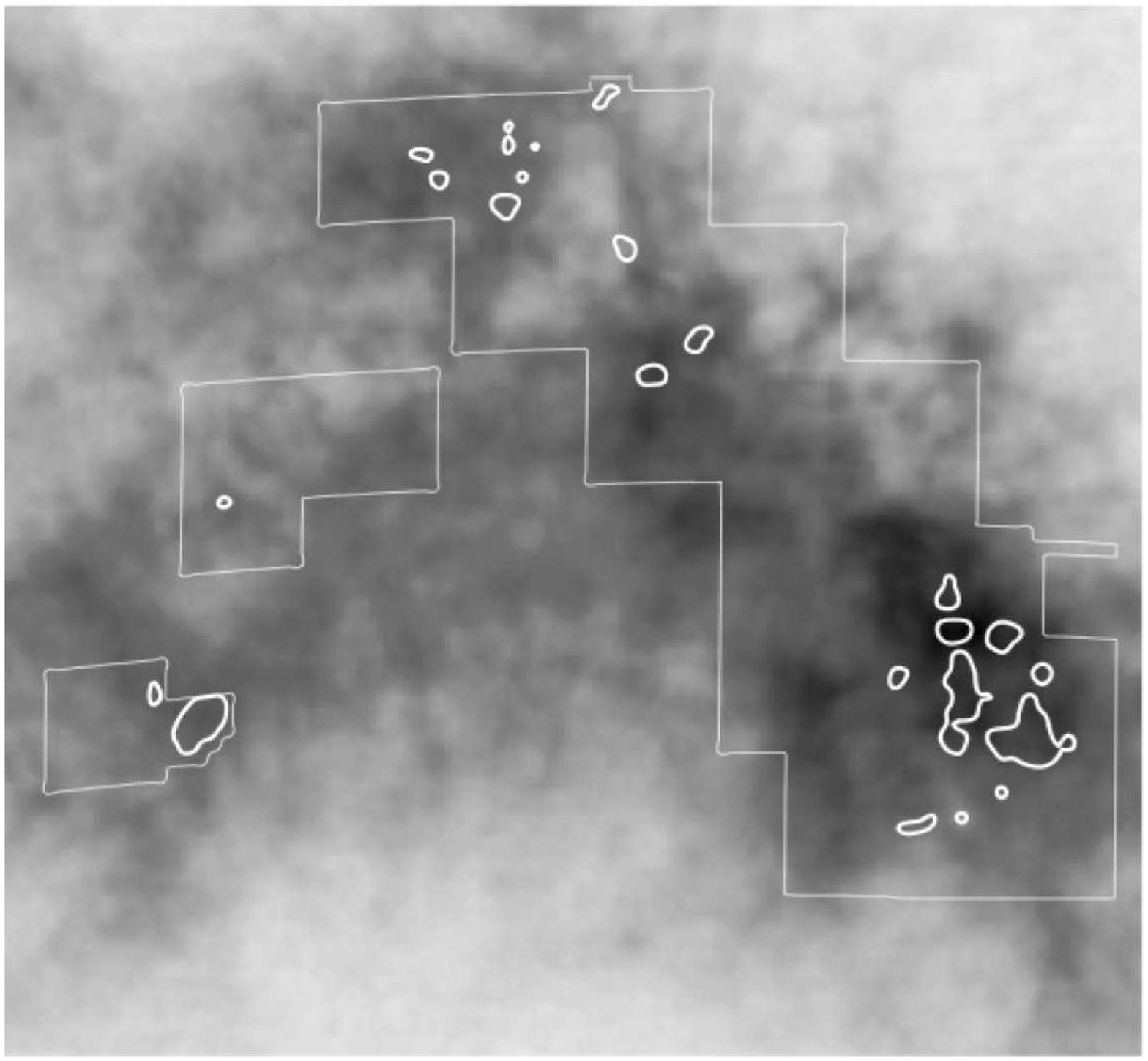}
\bigskip
\epsscale{2.25}
\plottwo{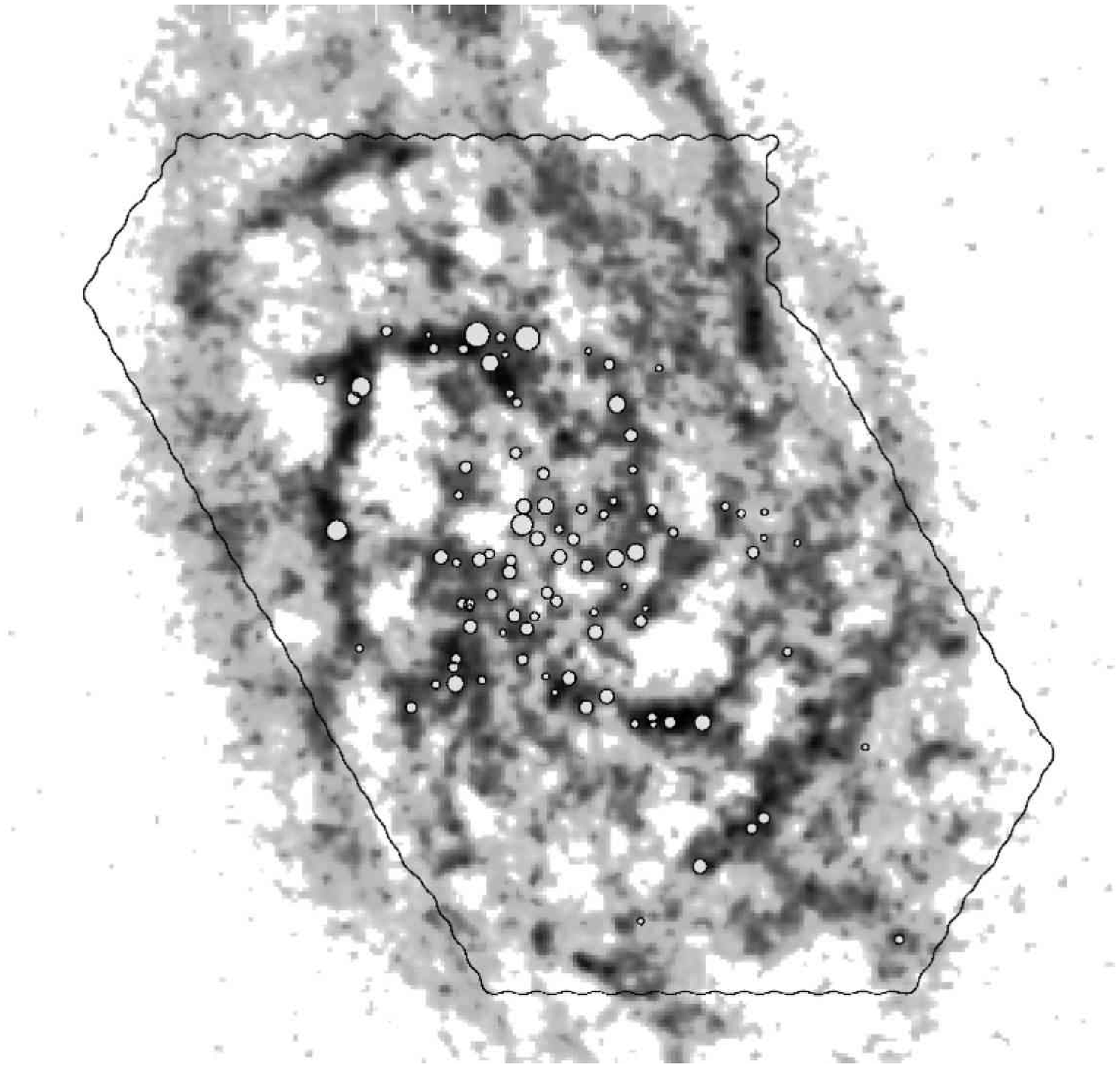}{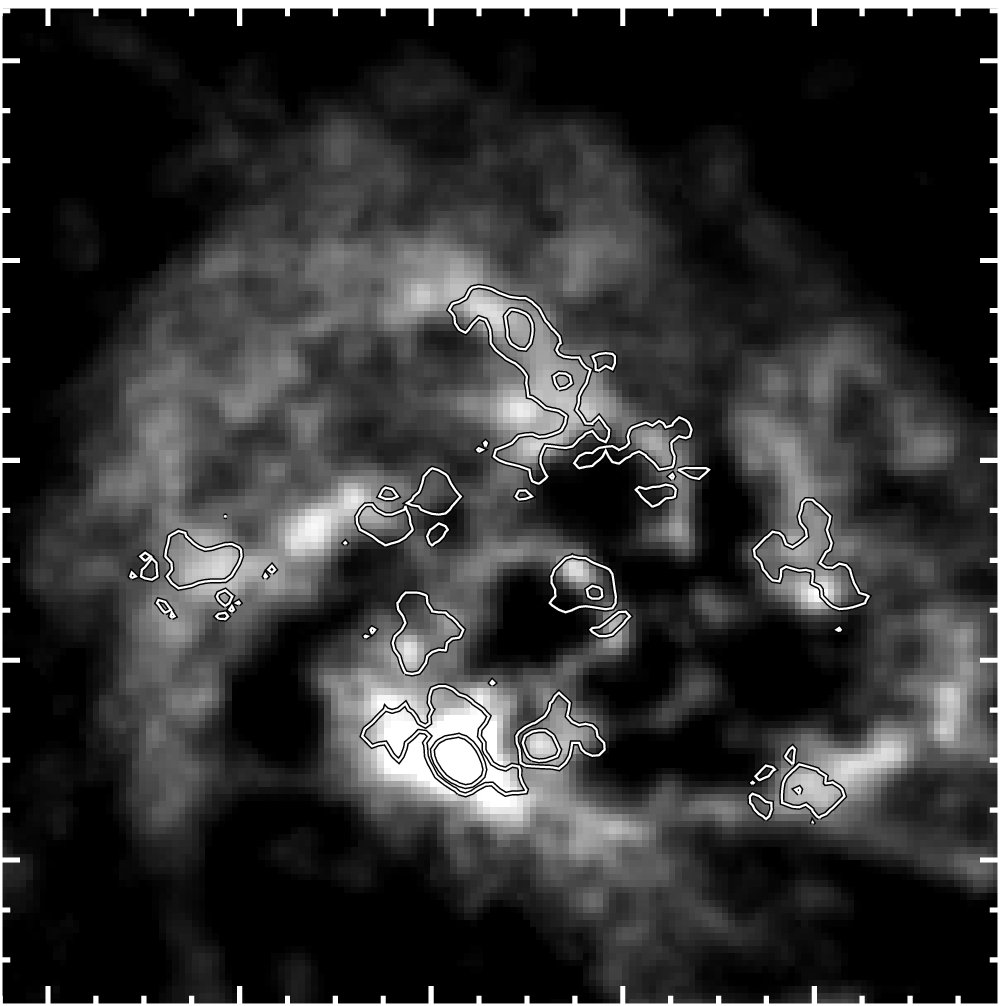}
\caption{\label{h1olays} CO emission overlayed on maps of
\ion{H}{1} emission for the LMC (top left), the SMC (top right), M33
(bottom left) and IC~10 (bottom right). The \ion{H}{1} maps are the
work of {\it Kim et al.}~(2003, LMC), {\it Stanimirovi\'c et
al.}~(1999, SMC), {\it Deul and van der Hulst} (1987, M33), and {\it
Wilcots and Miller} (1998, IC10).  Contours of the CO emission are
shown in each case except for M33 where the emission is indicated as
circles with area proportional to the flux.  Where appropriate, the
boundaries of the surveys are indicated.  CO emission is found
exclusively on bright filaments of atomic gas though not every bright
\ion{H}{1} filament has CO emission.}
\end{figure*}

The distribution of GMCs in these four galaxies shows little
correlation with old stars (see Figs.~\ref{lmcmap} and \ref{ic10map}).
The obvious correlations with H$\alpha$ (Figs.~\ref{lmcmap} and
\ref{m33map}) and young stellar clusters (Fig.~\ref{lmcmap}) are
expected since these trace the star formation that occurs within GMCs.
That the correlation is not perfect can be used to deduce information
about the evolution of the clouds (Section \ref{timescale}).  To examine
the relationship of GMCs to the remainder of the neutral ISM, we plot
the locations of CO emission on top of \ion{H}{1} maps of these four
galaxies in Fig.~\ref{h1olays}.  A strong correlation between the
atomic and molecular gas is immediately apparent.  Every GMC in each
of the galaxies is found on a bright filament or clump of \ion{H}{1},
but the reverse is not true: there are many bright filaments of
\ion{H}{1} without molecular gas.  In M33, the largest of the fully
mapped galaxies, the ratio of \ion{H}{1} to CO in the filaments in the
center of the galaxy is smaller than in the outer parts. In the LMC,
the CO is generally found at peaks of the
\ion{H}{1}, but most of the short filaments have no associated CO. In
the SMC, the \ion{H}{1} is so widespread that the CO clouds appear as
small, isolated clouds in a vast sea of \ion{H}{1}. Apparently,
\ion{H}{1} is a necessary but not a sufficient condition for the
formation of GMCs in these galaxies.

Figs.~\ref{lmcmap} -- \ref{ic10map} show that the molecular gas forms
from the \ion{H}{1}, rather than the \ion{H}{1} being a dissociation
product of the molecular clouds as some have advocated (e.g.,~{\it
Allen}, 2001). First, in all four galaxies the \ion{H}{1} is much more
widespread than the detected CO emission.  Thus, most of the
\ion{H}{1} cannot be dissociated H$_2$ without violating mass
conservation if the GMC lifetimes are as short as we derive in Section 6.
Second, there is no CO associated with most of the filaments in the
LMC, M33, and IC~10 and the column density of these CO-free filaments
is about the same as the column density of filaments which have CO
emission.  Because there is no transition in \ion{H}{1} properties at
radii where one observes CO, and the radii where it is absent, it is
difficult to imagine that two separate origins for the \ion{H}{1}
would produce a seamless transition.  Finally, the \ion{H}{1} in the
filaments between GMCs has the wrong geometry to be a dissociation
product; there is too much gas strung out along the filaments to have
come from dissociation of the molecular gas.

\bigskip
\subsection{Implications for GMC Formation}
\bigskip

What can the morphology of the atomic gas tell us about GMCs and their
formation?  All the \ion{H}{1} images are characterized by filamentary
structures that demarcate holes in the atomic distribution. In IC~10,
there is good evidence that some of the holes are evacuated by the
action of supernovae or stellar winds which sweep up the atomic gas
into the observed filamentary structure ({\it Wilcots and Miller},
1998).  In contrast, most of the large holes observed in the M33
\ion{H}{1} distribution are {\em not} likely to be caused by
supernovae. The large holes require about 10$^{53}$ ergs to evacuate,
but there are no obvious stellar clusters remaining at the center of
the holes.  Furthermore, x-ray emission is not concentrated in the
holes.  The large holes in M33 are thus likely to have a gravitational
or density-wave origin.  Small holes with $D < 200$~pc, on the other
hand, are found to be well correlated with OB associations ({\it Deul
and van der Hulst}, 1987); these tend to be concentrated toward the
center of the galaxy.

This leads to some qualitative conclusions about the formation of GMCs
and ultimately the star formation that occurs within them. Because the
CO forms from \ion{H}{1} filaments and not the other way around, it
is the filaments in a galaxy that must form first as precursors to the
GMCs.  In some of the galaxies, such as M33 and apparently in the LMC
and the SMC, most of the filaments are not associated with energetic
phenomena.  This clearly rules out the self-propagating star formation
picture that was promoted some years back by {\it Gerola and Seiden}
(1978) for most of our galaxies.  In their picture, GMC formation and
thus star formation propagates by means of supernovae that explode in
regions of a galaxy adjacent to a previous episode of star formation.
However, in IC~10, because there is evidence that some of the
\ion{H}{1} morphology may be the result of energetic events from
previous generations of stars, self-propagating star formation may be
a viable mechanism.  {\em The critical element of GMC formation across
all these systems appears to be the assembly of \ion{H}{1} filaments,
though the mechanism that collects the atomic gas appears to vary
across the systems.}

But why, then, do some filaments form GMCs and not others?  We argue in
Section 5, that it is the result of the pressure to which filaments are
subjected.

\begin{figure*}
\epsscale{1.0}
\plotone{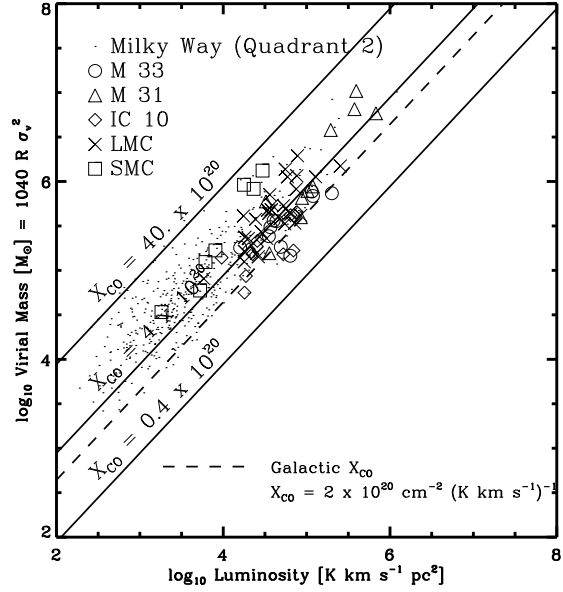}
\caption{\small Plot of the virial mass of the GMCs in our sample as a
function of luminosity. The value of \xco~from gamma-ray
investigations in the Milky Way ({\it Strong and Mattox}, 1996) is
shown by the dashed line.  The plot shows that while there are some
differences in \xco~from galaxy to galaxy, except for the SMC, a value
of $\xco = 4 \times 10^{20}$ cm$^{-2} (\Kkmpers)^{-1}$ can be used for
all of the other galaxies to a reasonable degree of
approximation. \label{xcofig}}
\end{figure*}

\begin{deluxetable}{l c c}
\tabletypesize{\large}
\tablewidth{0pt}
\tablecolumns{5}
\tablecaption{\label{XCOTAB} $X_{CO}$ Across the Local Group}

\tablehead{ \colhead{Galaxy} & \colhead{Mean $X_{CO}$} &
\colhead{Scatter in $X_{CO}$\tablenotemark{a}} \\
\colhead{} & \colhead{$\times 10^{20}$ \xcounits} &
\colhead{$\times 10^{20}$ \xcounits} }

\startdata

SMC & $13.5 \pm 2.6$ &  $2.2$ \\
LMC & $5.4 \pm 0.5$ &  $1.7$ \\
IC10 & $2.6 \pm 0.5$ &  $2.2$ \\
M33 & $3.0 \pm 0.4$ &  $1.5$  \\
M31 & $ 5.6 \pm 1.1$ &  $2.7$ \\
Quad 2\tablenotemark{b} & $6.6 \pm 0.6$ &  $2.0$ \\
Local Group\tablenotemark{c} & $5.4 \pm 0.5$ & $2.0$ \\
\enddata
\tablenotetext{a}{Scatter is a factor based on median absolute deviation of the log.}
\tablenotetext{b}{Clouds with luminosities corresponding to $M_{Lum}
  \geq 5 \times 10^4$ M$_{\odot}$ (for $X_{CO} = 2\times10^{20}$)}
\tablenotetext{c}{Excluding Milky Way.}
\end{deluxetable}

%One feature that all of the galaxies have in common is that all of the
%GMCs are clearly associated with \ion{H}{1} filaments or local
%\ion{H}{1} density peaks.  However, not all of the filaments have GMCs
%associated with them, even where \ion{H}{1} surface densities are the
%same as those containing GMCs.  In M33, the largest of the fully
%mapped galaxies, the ratio of \ion{H}{1} to CO in the filaments in the
%center of the galaxy is smaller than in the outer parts. In the LMC,
%the CO is generally found at peaks of the \ion{H}{1}, and most of the
%short filaments have no associated CO. In the SMC, the \ion{H}{1} is
%so widespread that the CO clouds appear as small, isolated clouds in a
%vast sea of \ion{H}{1}. Apparently, \ion{H}{1} is a necessary, but not
%a sufficient condition for the formation of GMCs in these galaxies.

%We note that in all four galaxies, the \ion{H}{1} is much more
%widespread than the CO emission, and extends far beyond the region of
%detectable CO.  Clearly, the CO must form from the \ion{H}{1}, rather
%than the \ion{H}{1} being a dissociation product of the molecular
%clouds as some have advocated (e.g.,~Allen 2001).

\bigskip
\section{\textbf{MOLECULAR CLOUD PROPERTIES}}
\bigskip

Our main goal in this section is to compare the properties of GMCs
made with different telescopes, resolutions, and sensitivities.  We
use GMC catalogs from the studies of the four galaxies listed above,
and we supplement our work with a sample of GMCs in M31 ({\it
Rosolowsky}, 2006) as well as a compilation of molecular clouds in the
outer Milky Way as observed by {\it Dame et al.}~(2001) and cataloged
in {\it Rosolowsky and Leroy} (2006).
%  In \S 5 we also discuss the
%GMCs found in the center of M64 (Rosolowsky and Blitz 2005).

To aid in the systematic comparison of cloud properties, {\it
Rosolowsky and Leroy} (2006, hereafter RL06) have recently published a
method for minimizing the biases that plague such comparisons.  For
example, measurement of the cloud radius depends on the sensitivity of
the measurements, and RL06 suggest a robust method to extrapolate to
the expected radius in the limit of infinite sensitivity.  They also
suggest a method to correct cloud sizes for beam convolution, which
has been ignored in many previous studies of extragalactic clouds.  We
use the RL06 extrapolated moment method on all of the data used in
this paper since it is least affected by relatively poor
signal-to-noise and resolution effects.  We have also applied the RL06
methodology to the outer Milky Way data of {\it Dame et al.}~(2001)
rather than relying on published properties (e.g.,~{\it Heyer et
al.},~2001).  It is for this reason that we have not included the
cloud properties of {\it Solomon et al.}~(1987) in our plots, but we
do make comparisons to their work at the end of this section. Except
where noted, we consider only clouds that are well-resolved by the
telescope beam; the GMCs must have angular diameters at least twice
that of the beam used to observe them.

Are we seeing single or multiple objects in the beam?  The issue of
velocity blending of multiple clouds in the beam is much less of an
issue in extragalactic observations than in the Galactic case, where
the overwhelming majority of GMCs are observed only in the Galactic
plane.  Extragalactic observations of all but the most highly inclined
galaxies do not suffer from this problem and as can be seen in 
Figs.~1 -- 4, the clouds are, in general, spatially well separated, ensuring
that we are almost always seeing only a single GMC along the line of
sight.

One of the long debated questions related to GMCs is: how does
metallicity affect the value of \xco, the conversion factor from CO
line strength to H$_2$ column density? Fig.~\ref{xcofig} is a plot of
the virial mass of the GMCs as a function of CO luminosity.  Diagonal
lines are lines of constant \xco. A compilation of
\xco~values is given in Table 2.  We note first that most of the
points lie above the dashed line that indicates the value determined
from gamma-rays in the Milky Way ({\it Strong and Mattox}, 1996).  A value of
$\xco = 4\times 10^{20}$ cm$^{-2} (\Kkmpers)^{-1}$ would allow virial
masses to be derived to within about a factor of two for all of the
GMCs in our sample, with the clouds in the SMC and the outer Galaxy
requiring a somewhat higher value.

Note, however, that the SMC clouds are systematically higher in this
plot than the GMCs for any other galaxy, and that the GMCs in IC~10
are systematically a bit lower.  Solving for \xco~in the SMC, gives a
value of 13.5 $\times 10^{20}$ cm$^{-2} (\Kkmpers)^{-1}$, more than a
factor of 3 above the mean.  In contrast, IC~10 yields $\xco =2 \times
10^{20}$ cm$^{-2} (\Kkmpers)^{-1}$.  Surprisingly, the galaxies differ
in metallicity from one another only by a factor of two, and both are
much less than solar.  In M33, the metallicity decreases by almost an
order of magnitude from the center out ({\it Henry and Howard}, 1995),
but {\it Rosolowsky et al.}~(2003) find no change in \xco~with
radius.  Although metallicity may be a factor in determining \xco~in
different galaxies, there is no clear trend with metallicity alone --
other factors appear to be as important as the metallicity in
determining
\xco.

 \begin{figure*}
\epsscale{2.0}
 \plottwo{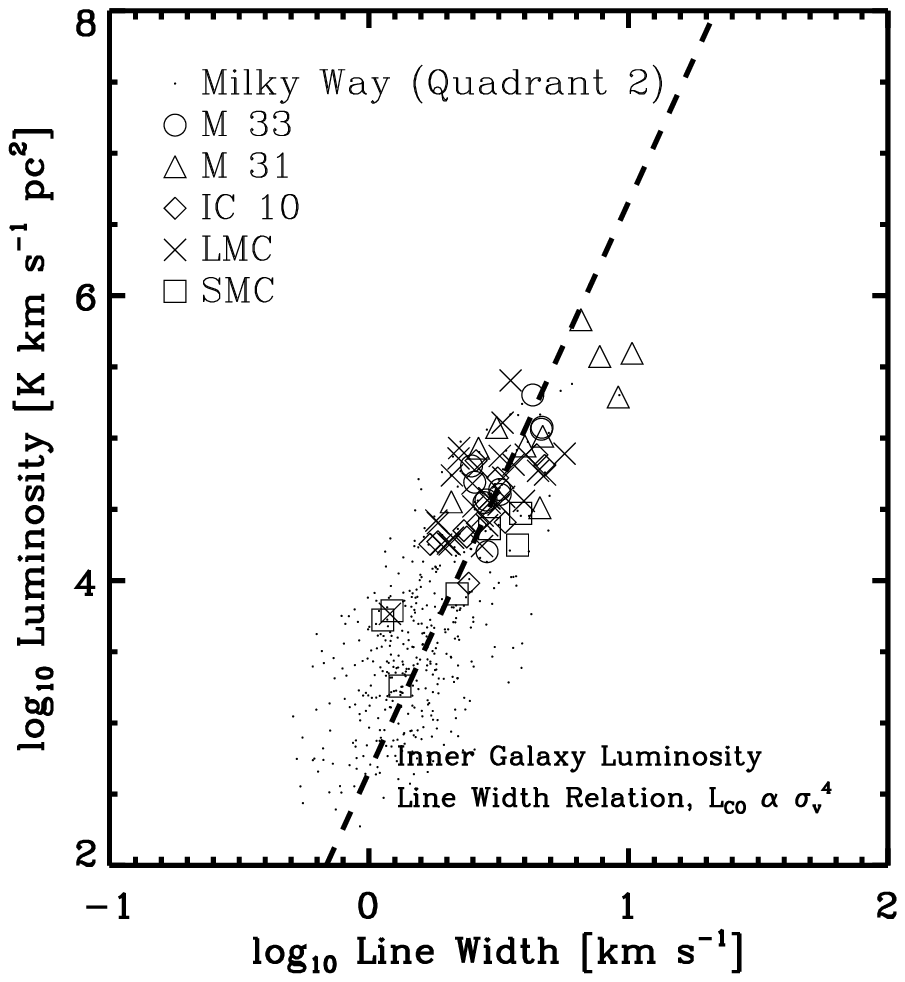}{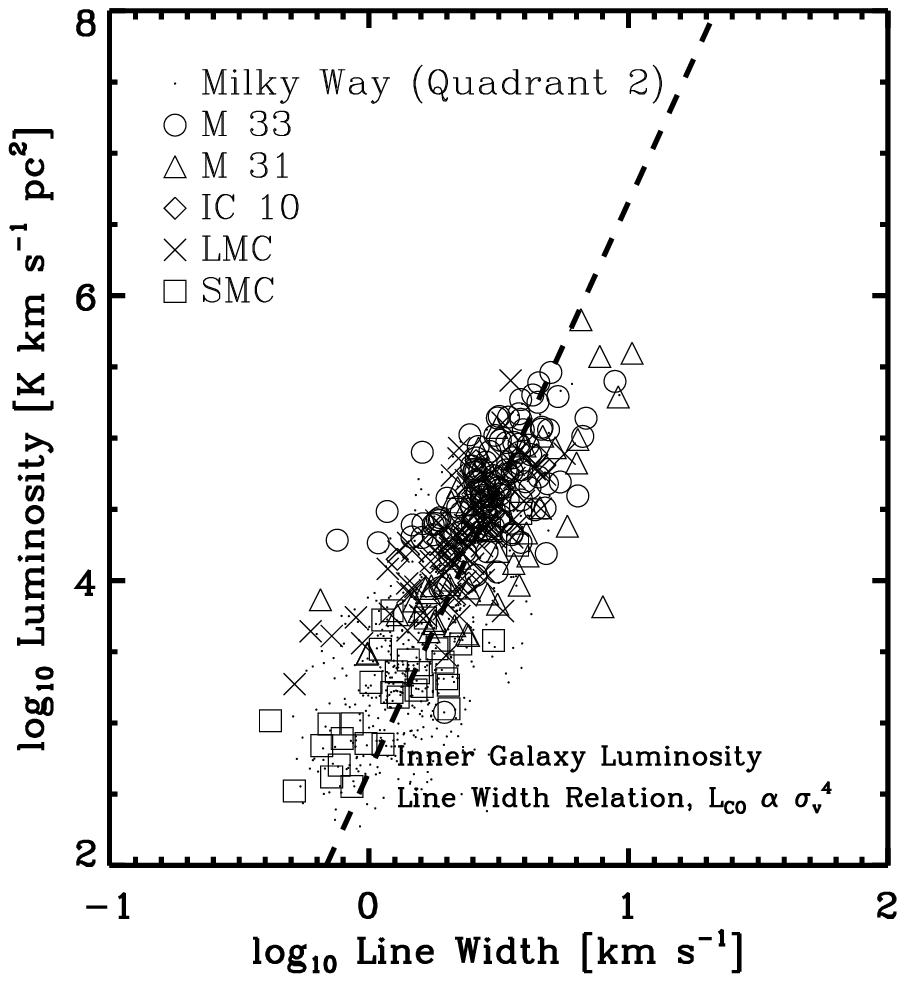}
\caption{\small {\it (Left)} Luminosity vs. Line width plot for all of
 the resolved clouds in our survey.  The dashed line, $L_{\mathrm{CO}}
 \propto {\sigma_v}^4$, with a single constant of proportionality is a
 good representation of the data. {\it (Right)} The same as the
 left-hand panel but including the unresolved clouds in our sample.
 The dashed line remains a good representation of the data even with
 much more data included.\label{ldv} }
\end{figure*}

The discrepancy between the Galactic gamma-ray value of 2 $\times
10^{20}$ cm$^{-2} (\Kkmpers)^{-1}$ and the virial value we derive here
is not necessarily a problem.  Taken at face value, it may be telling
us is that the GMCs are not in virial equilibrium, but are nearly
gravitationally neutral: the overall potential energy is
equal to the kinetic energy.  The gamma-ray value of \xco~is
independent of the dynamical state of the cloud, thus, uncertainties
about the self-gravity of GMCs do not come into play.  Since GMCs do
not look as if they are in virial equilibrium (they are highly
filamentary structures and do not appear to be strongly centrally
concentrated), these two different values of \xco~are consistent if
the clouds are only marginally self-gravitating.

\begin{figure*}
\epsscale{2.2}
\plottwo{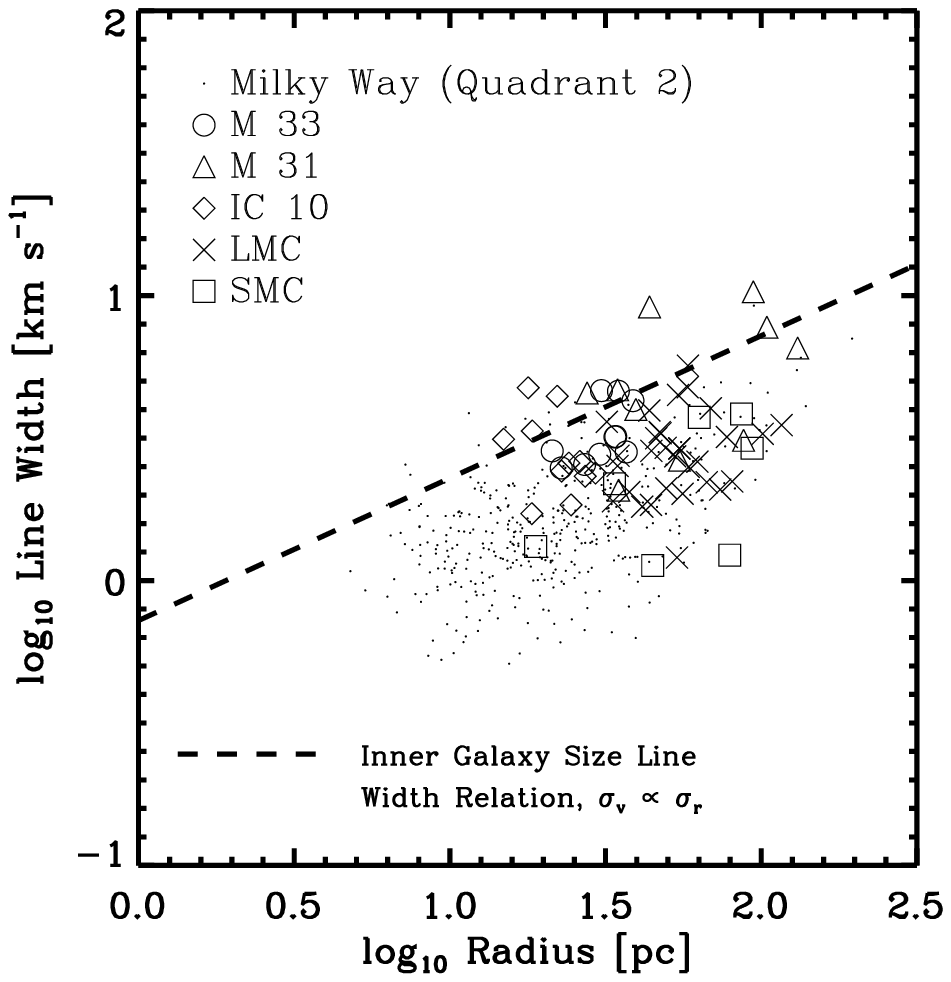}{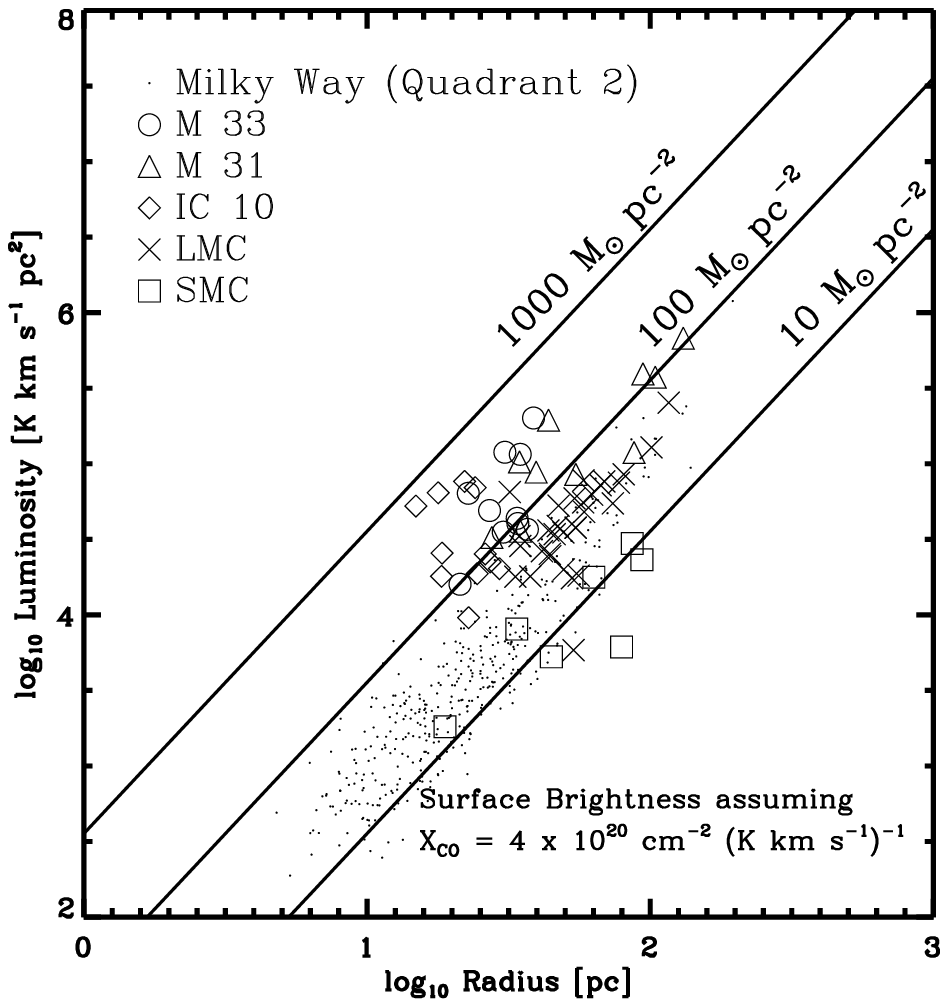}
\caption{\small {\it (left)} Line width-size relation for the GMCs in
our sample. The dashed line is the relation found for the GMCs in the
inner Milky Way, showing a clear offset from the extragalactic GMCs.
{\it (Right)}  Luminosity vs. Radius relation for the GMCs in our
sample. Solid lines are lines of constant surface density assuming
$\xco = 4 \times 10^{20}$ cm$^{-2} (\Kkmpers)^{-1}$. The
galaxies show clear differences in CO luminosity for a given cloud
radius. \label{larson}}  
\end{figure*}

The left-hand panel of Fig.~\ref{ldv} is a plot of the CO luminosity
of GMCs as a function of line width. It may be thought of as a plot of
\h2 mass vs. line width for a single, but undetermined, value of \xco.
The dashed line, is the relation $L_{\mathrm{CO}} \propto
{\sigma_v}^4$, is not a fit, but is a good representation of the data
for both the five external galaxies in our sample as well as for the
outer Milky Way. The scatter in the relationship is 0.5 dex, or a
factor of 3 over three orders of magnitude in luminosity.  If the GMCs
are self-gravitating, then they obey
\begin{equation}
M = 5 R {\sigma_v}^2/(\alpha G)
\end{equation}
where $\alpha$ is a constant of order unity. Provided the CO luminosity is
proportional to the mass of a GMC, the plot shows that $M$(\h2)~$\propto
{\sigma_v}^4$; thus

\begin{equation}
\sigma_v \propto R^{0.5}~~~{\rm and}~~~M/R^2 = {\rm constant}.
\end{equation} 
These two relations are shown on the left- and right-hand sides of
Fig.~\ref{larson} respectively.

The advantage of a luminosity-line width plot, especially for
extragalactic studies is that one need not resolve the individual
clouds, since the luminosity, and by implication, the mass, is
independent of resolution.  One need only be sure that individual GMCs
are isolated in the beam.  The right-hand panel in Fig.~\ref{ldv}
shows all of the individual clouds identified in the galaxy surveys,
most of which are unresolved.  We see that the clouds populate the
same $L_{\mathrm{CO}} \propto {\sigma_v}^4$ line as in the left-hand panel.
This plot demonstrates probably better than any other that the GMCs in
our sample are much more alike than they are different.

The left-hand side of Fig.~\ref{larson} is the size-line width
relation for the GMCs in our sample.  The dashed line is the size-line
width relation for GMCs in the inner region of the Milky Way from {\it
Solomon et al.}~(1987).  First, we note that the correlation for the
extragalactic clouds is very weak.  However, if we add the outer
Galaxy clouds, the correlation does seem to be consistent with a power
law relation $\sigma_v \propto R^{0.5}$. However, there is a clear
offset from the relation determined for the inner Galaxy (dashed line,
{\it Solomon et al.},~1987).  At least part of this offset can be attributed
to differences in the methods used to measure cloud properties.  The
sense of the offset is that for a given cloud radius, inner Milky Way
clouds have larger line widths. This may be partially due to the
relatively high value of $T_A$ used by {\it Solomon et al.}~(1987) to
define the cloud radius, implying that the clouds might be inferred 
to be smaller for
a given value of $\sigma_v$.

But part of the offset may also be real.  We see that there is a clear
separation of the clouds by galaxy in the plot.  The IC~10 clouds lie
to the left of the diagram, while the LMC clouds lie to the right.
The SMC clouds tend to lie at the bottom of the group. The apparently
weak correlation of extragalactic clouds is probably due to the small
dynamic range in the plot compared to the measurement error in the
cloud properties; the rms scatter in Fig.~\ref{larson} (left) is
only 0.2 dex, or less than a factor of two.  We therefore conclude
that the GMCs in our sample are consistent with a power law relation
$\sigma_v \propto {R}^{0.5}$.  There are, however, real differences in
the coefficient of proportionality, and this gives rise to some of the
scatter in the relationship.  The size-line width relationship arises
from the turbulent nature of the molecular gas motions.  Differences
in the constant of proportionality imply variations in the
normalization of the turbulent
motions of GMCs in different galaxies, independent of cloud
luminosity.

\begin{figure*}
\epsscale{1.0}
\plotone{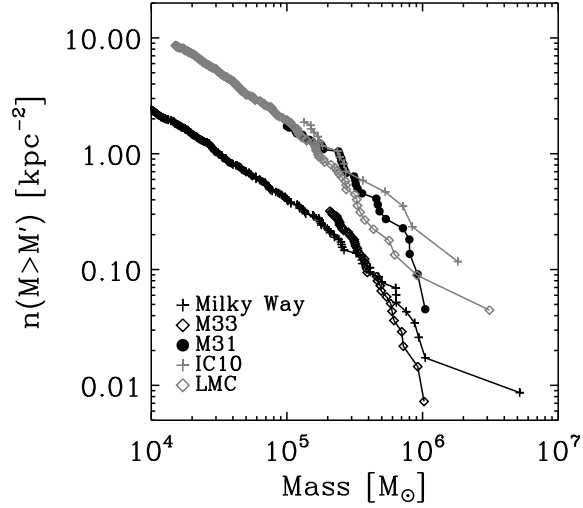}
\caption{\small Cumulative mass distribution for the Galaxies in our sample.
The mass distributions have been normalized by the area surveyed in
each galaxy. In this figure, we use all clouds above the completeness
limits in each survey, whether or not the clouds are resolved. All of
the galaxies look similar except for M33 which has a steeper mass
spectrum than the others.\label{massspec}}
\end{figure*}

\begin{deluxetable}{l  c }
\tabletypesize{\large}
\tablewidth{0pt}
\tablecolumns{5}
\tablecaption{\label{massspec} Mass Distributions of the 5 Galaxies}

\tablehead{ \colhead{Galaxy} &  \colhead{Index}}

\startdata

LMC &  $ -1.74 \pm 0.08$ \\
IC10 & $ -1.74 \pm 0.41$  \\
M33 &  $ -2.49 \pm 0.48$ \\
M31 &  $ -1.55 \pm 0.20$ \\
Outer MW & $ -1.71 \pm 0.06$ \\

\enddata
\end{deluxetable}

These conclusions help to explain Fig.~\ref{larson} (right), which
is a plot of luminosity vs. radius.  Assuming that luminosity is
proportional to mass, at least within a single galaxy, we can plot
lines of constant surface brightness.  After all, Fig.~\ref{xcofig}
suggests that the clouds have a constant surface brightness.  In fact,
it appears that for a given galaxy, the individual GMCs are strung out
along lines of constant surface density, but with each galaxy lying on
a different line.  The SMC clouds, for example, have a mean surface
density of ~10 $M_\sun$ pc$^{-2}$, but the IC~10 clouds have a mean
surface density $>$ 100 $M_\sun$ pc$^{-2}$.  A direct interpretation
of Fig.~\ref{larson} (right) implies that for a given radius, the
SMC clouds are less luminous than the rest, and the IC~10 clouds
are more luminous.  Another way of saying this is that for a given
cloud luminosity, the SMC clouds are larger, as are the LMC clouds,
only less so.  This difference disappears, for the most part, if we
consider the mass surface density rather than the surface brightness.
In that case one must multiply the luminosity of the GMCs in each galaxy by its
appropriate value of \xco. When that is done, the
difference in the mean surface density from galaxy to galaxy 
is less than a factor of two.

In Fig.~\ref{ldv} we see that the GMCs in the SMC are well-separated
from the GMCs in M31, implying that the median luminosity of the two
sets of clouds is different by nearly two orders of magnitude. The
differences due to \xco~ are only a factor of about 4; but is 
the distribution of GMC masses in the two systems really
different? There are not enough clouds to measure a mass spectrum in the
SMC, but Fig.~\ref{massspec} shows the mass distribution of molecular
clouds normalized to the survey area for the other five galaxies.
Power-law fits to the masses of all cataloged molecular clouds above
the completeness limit give the index of the mass distributions listed
in Table 3.  All of the galaxies have remarkably similar mass
distributions except for M33, which is much steeper than the others.
In addition, the mass distributions in M31 and the LMC show a
truncation at high mass similar to that found in the inner Milky Way
(e.g.,~{\it Williams and McKee}, 1997) suggesting that there is a
characteristic cloud mass in these systems.  In addition, {\it
Engargiola et al.}~(2003) also argue for a characteristic cloud mass 
in M33 but it is not at the high mass end, as it is for the LMC
and M31; rather it has a value of 4 -- 6 $\times 10^4$ M$_\sun$.  
The variation in the mass distributions is unexplained and
may offer an avenue to understanding differences in star formation
rates between galaxies.

\bigskip 
\section{\textbf {THE ROLE OF HYDROSTATIC PRESSURE}}
\bigskip
\label{pressure}

A number of authors have speculated on the role that hydrostatic
pressure plays in the formation of molecular clouds in the centers of
galaxies ({\it Helfer and Blitz}, 1993; {\it Spergel and Blitz},
1992), and galactic disks ({\it Elmegreen}, 1993; {\it Wong and Blitz},
2002; among others).  {\it Blitz and Rosolowsky} (2004) showed that if
hydrostatic pressure is the only parameter governing the molecular gas
fraction in galaxies, then one predicts that the location where the
ratio of molecular to atomic gas is unity occurs at constant {\it
stellar} surface density.  They probed this prediction and found that
the constancy holds to within 40\% for 30 nearby galaxies.

The functional form of the relationship between hydrostatic pressure
and molecular gas fraction has recently been investigated by {\it
Blitz and Rosolowsky} (2006) for 14 galaxies covering 3 orders of
magnitude in pressure.  Hydrostatic pressure is determined by

\begin{equation}
P_{hydro} =  0.84 (G \Sigma_*)^{0.5}\Sigma_g \frac {v_g} {(h_*)^{0.5}}. \
\label{approxpressure}
\end{equation}

\begin{figure}[bt]
\epsscale{1.0}
\plotone{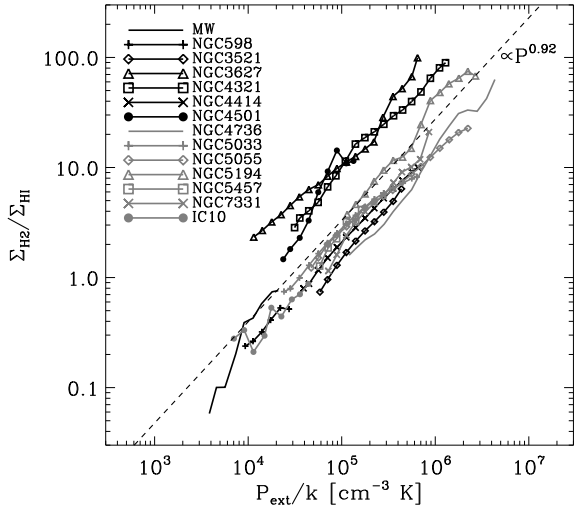}
\caption{\small Plot of the ratio of molecular to atomic surface
density as a function of hydrostatic pressure for 14 galaxies.  The
plot covers 3 magnitudes in pressure and molecular fraction. \label{press}}  
\end{figure}

\noindent The quantities ${v_g}$, the gas velocity dispersion, and
$h_*$, the stellar scale height, vary by less than a factor of two
both within and among galaxies ({\it van der Kruit and Searle}, 1981a,b;
{\it Kregel et al.}, 2002). The quantities $\Sigma_*$,
the stellar surface density, and $\Sigma_g$, the gas surface density,
can be obtained from observations. The results for 14 galaxies is
given in Fig.~\ref{press}.

The figure shows that the galaxies all have similar slopes for the
relationship: $\Sigma_{\mathrm{H2}}/\Sigma_{\mathrm{HI}} \propto
P^{0.92}$, very nearly linear.  Moreover, except for three galaxies,
NGC 3627, NGC 4321, and NGC 4501, all have the same constant of
proportionality.  The three exceptional galaxies all are interacting
with their environments and may be subject to additional pressure
forces.  It is important to point out that we expect this pressure
relation to break down at some lower scale no smaller than the scale
of a typical GMC, $\sim$50 pc.  However, on the scale of the pressure
scale height, typically a few hundred parsecs, the pressure should be
more or less constant both vertically and in the plane of a galaxy.

The two axes in Fig.~ 10 are not completely independent; both are
proportional to $\Sigma_{\mathrm{H2}}$.  However, each axis is also
dependent on other quantities such as $\Sigma_{\mathrm{HI}}$ and
$\Sigma_*$.  Since $\Sigma_*$ varies by a larger amount in a given
galaxy than $\Sigma_{H_2}$, because $\Sigma_{\mathrm{HI}}$ dominates
at low pressures ($P/k < 10^5~\mbox{cm}^{-3}~\mbox{K}$) and
because both axes have different dependencies on
$\Sigma_{\mathrm{H2}}$, the constancy of the slopes and the agreement
of the intercepts cannot be driven by the common appearance of
$\Sigma_{\mathrm{H2}}$ on each axis. A more detailed discussion of
this point is given in {\it Blitz and Rosolowsky} (2006).

% values of  $\Sigma_{H_2}/\Sigma_{\ion{H}{1}} > 5$, 
As of this writing we do not know how the LMC and the SMC fit into
this picture; no good map giving the stellar surface density for these
objects is currently available.  Although we do not know the stellar
scale heights for these galaxies, because of the weak dependence on
$h_*$ in Equation 3, this ignorance should not be much of a
difficulty.  The results for the SMC are particularly interesting
because of its low metallicity and low dust-to-gas ratio ({\it Koorneef},
1984; {\it Stanimirovic et al.},~2000). Since the extinction in the UV is
significantly smaller than in other galaxies, one might expect higher
pressures to be necessary to achieve the same fraction of molecular
gas in the SMC, though care must be taken since CO may be compromised
as a mass tracer in such environments.

\medskip

The following picture for the formation of molecular clouds in
galaxies is, then, suggested by the observations.  Density waves or
some other process collects the atomic gas into filamentary
structures.  This process may be the result of energetic events, as is
thought to be the case for IC~10, or dynamical processes, as is
primarily the case for M33.  Depending on how much gas is collected,
and where in the gravitational potential of the galaxy the gas is
located, a fraction of the atomic hydrogen is turned in molecular gas.
In very gas-rich, high pressure regions near galactic centers, this
conversion is nearly complete.  But some other process, perhaps
instabilities, collects the gas into clouds.  Whether this is done
prior to the formation of GMCs, or after is not clear.

\bigskip
\section{\textbf {GMCS IN STARBURST GALAXIES}}
\bigskip

In many galaxies the average surface density of molecular gas is much
greater than the surface densities of individual GMCs shown in
Fig.~\ref{larson} ({\it right}; {\it Helfer et al.},~2003).  These
regions of high surface density can be as much as a kiloparsec in
extent.  Indeed, about half of the galaxies in the BIMA SONG survey
({\it Helfer et al.},~2003) have central surface densities in excess
of 100 $M_\sun$~pc$^{-2}$.  Moreover, regions with high surface
densities of molecular gas are invariably associated with dramatically
enhanced star formation rates ({\it Kennicutt}, 1998).  In regions of
such high surface density, are there even individual, identifiable
GMCs?  If so, do they obey the same relations shown in
Figs.~\ref{xcofig} --
\ref{larson}?

Several recent studies have begun to attack these questions.  The
only such molecule-rich region in the Local Group is in the vicinity
of the Galactic Center where cloud properties were analyzed by {\it
Oka et al.}~(2001).  They found that clouds in the Galactic center
were smaller, denser and had larger line widths than the GMCs in the
Galactic disk.  For targets beyond the Local Group, achieving the
requisite spatial resolution to study individual GMCs requires
significant effort.  To date, only a few extragalactic, molecule-rich
regions have been studied.  {\it Keto et al.}~(2005) show clouds in
M82 to be roughly in virial equilibrium.  At the high surface
densities of molecular gas observed in M82, this requires clouds to be
smaller and denser than those found in the Galactic disk.  Similarly,
{\it Rosolowsky and Blitz} (2005) observed the inner region of the
Galaxy M64, which has a surface density of $\sim$ 100 $M_\sun$
pc$^{-2}$ over the inner 300 pc of the galaxy.  They found 25 GMCs
with densities 2.5 times higher, and are 10 times more massive, on average,
than typical disk GMCs. This conclusion is quite robust against
differences in cloud decomposition because if some of the clouds they
identify are in fact blends of smaller clouds, then the derived
densities are lower limits, reinforcing their conclusions.  In M64,
{\it Rosolowsky and Blitz} (2005) examine many of the relationships
shown in Figs.~ \ref{ldv} and \ref{larson} and find that all are
significantly different.

\subsection{GMC Formation in Galactic Centers}

The peak \htwo~surface density in the central 1 kpc of M64 is about 20
times the \hi~surface density ({\it Braun et al.}, 1994; {\it
Rosolowsky and Blitz}, 2005), which is typical of many galaxies ({\it
Helfer et al.}, 2003).  In such regions, the formation of GMCs cannot
take place by first collecting atomic hydrogen into filaments and then
turning the gas molecular.  If the gas is cycled between the atomic
and molecular phases, as is required by the presence of \ion{H}{2}
regions in the central regions of M64, then continuity
requires that the amount of time that the gas remains in each phase is
roughly equal to the ratio of surface densities at each particular
radius.  Thus, gas ionized by the O stars must quickly return to the
molecular phase, which is catalyzed by the very large pressures in the
central region (Section 4).  More than likely, the GMCs are formed and
destroyed without substantially leaving the molecular phase, unlike
what happens in the disks.  Indeed, {\it Rosolowsky and Blitz} (2005)
present evidence for a diffuse molecular component that is not bound
into GMCs.  Thus it seems likely that, as in galactic disks, the
formation of structure (filaments?) in galactic nuclei occurs before
the formation of the GMCs.  The gas, though, is largely molecular
prior to the formation of the clouds.

Measuring the properties of individual GMCs in more distant
molecule-rich galaxies will rely upon future improvements in angular
resolution and sensitivity.  At present, some information can be
gleaned from single-dish spectra of the regions in multiple tracers of
molecular gas.  The observations of {\it Gao and Solomon} (2004) and
{\it Narayanan et al.}~(2005) show that the star formation rate is
linearly proportional to the mass of molecular gas found at high
densities ($\ge 10^5~\mbox{cm}^{-3}$), and that the fraction of dense
gas increases with the amount of molecular mass in the system.  Since
the fraction of molecular mass found at high densities is relatively
small in Galactic GMCs, this implies there are substantial differences
in GMC properties in these starburst systems.

\begin{figure*}[ht!]
\epsscale{1.55}
\plotone{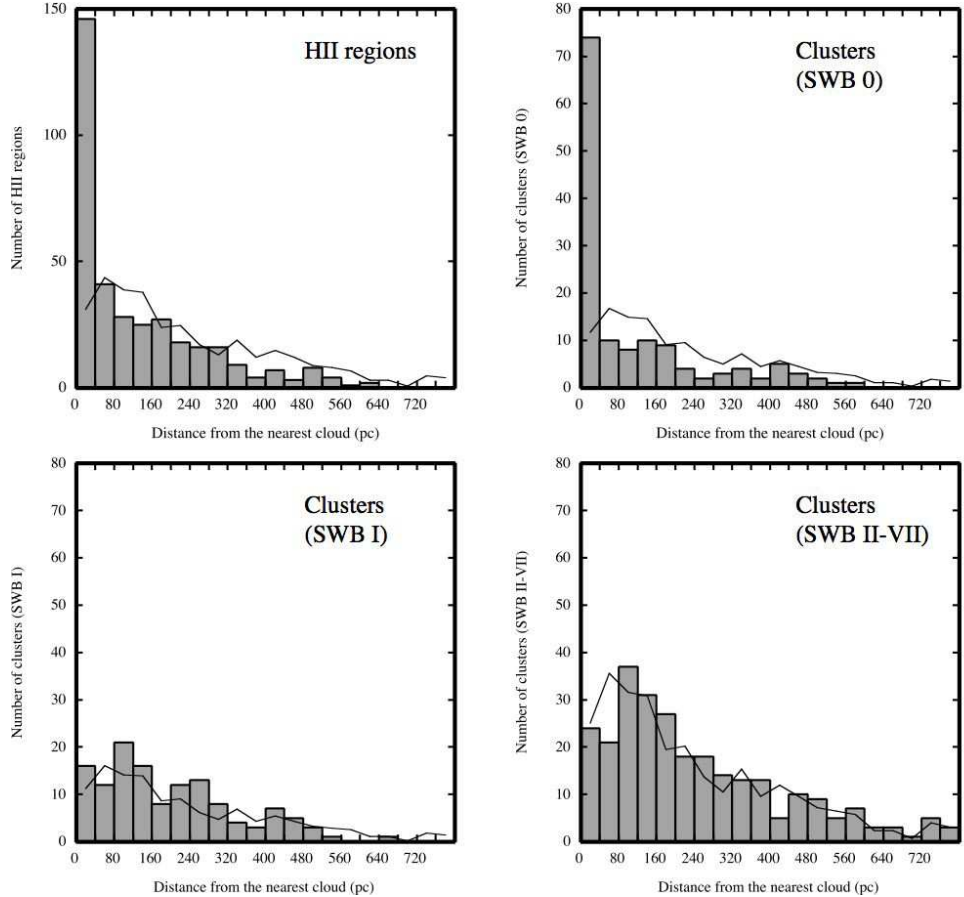}
\caption{\small Histograms of the projected separation from the \ion{H}{1}I regions 
({\it Top left} {\it Davies et al.}~1976) and clusters cataloged by 
{\it Bica et al.}~(1996) to the neareset CO emission; 
({\it Top right}) clusters with $\tau < 10$ Myr (SWB 0),
({\it Bottom left}) clusters with  10 Myr $< \tau < 30$ Myr (SWB I), and
({\it Bottom right}) clusters with  30 Myr $< \tau $ (SWB II - VII), respectively.
The lines represent the frequency distribution expected if the same number
of the clusters are distributed at random in the observed area.
\label{histo}}
\end{figure*}

\bigskip
\section{\textbf {STAR FORMATION IN EXTRAGALACTIC GMCs}}
\bigskip
\label{timescale}

The evolution of GMCs substantially influences the evolution of
galaxies. In particular, star formation in GMCs is a central event
that affects galactic structure, energetics, and chemistry.  A detailed
understanding of star formation is therefore an important step for a better
understanding of galaxy evolution.

\bigskip
\noindent
\subsection{\textbf{Identification of Star Formation}}
\bigskip

In Galactic molecular clouds, we are able to study the formation of
stars from high mass to low mass including even brown dwarfs.  In all
external galaxies, even those in the Local Group, such studies are
limited to only the highest mass stars as a result of limited
sensitivity. It is nonetheless worthwhile to learn how high-mass stars
form in GMCs because high-mass stars impart the highest energies to
the ISM via UV photons, stellar winds, and supernova explosions.

Young, high-mass stars are apparent at optical/radio
wavelengths as the brightest members of stellar clusters or associations
or by the H$\alpha$ and radio continuum emission from \ion{H}{2}
regions. The positional coincidence between these signposts of star
formation and GMCs is the most common method of identifying the star
formation associated with individual clouds.  Such associations can be
made with reasonable confidence when the source density is small enough
that confusion is not important.  When confusion becomes significant, however,
conclusions can only be drawn by either making more careful
comparisons at higher angular resolution should or by adopting a
statistical approach.

\bigskip
\noindent
\subsection{\textbf{The Large Magellanic Cloud}}
\bigskip

The most complete datasets for young stars are available for the LMC,
which has a distance of 50 kpc.  They include catalogs of clusters and
associations (e.g., {\it Bica et al.}, 1996) and of optical and radio
\ion{H}{2} regions ({\it Henize}, 1956, {\it Davies et al.}, 1976, 
{\it Kennicutt and Hodge}, 1986, {\it Filipovic et al.}, 1998). The
colors of the stellar clusters are studied in detail at four optical
wavelengths and are classified into an age sequence from SWB0 to SWB
VII, where SWB0 is the youngest with an age of less than 10 Myr, SWB I
in a range 10--30 Myr, and so on ({\it Bica et al.}, 1996).  The
sensitivity limit of the published catalogs of star clusters is 14.5
mag ($V$); it is not straightforward to convert this into the number
of stars since a stellar mass function must be assumed. The datasets
of \ion{H}{2} regions have a detection limit in H${\alpha}$ flux of
$10^{-12}$ ergs cm$^{-2}$ s$^{-1}$, and the radio sensitivity limit at
5 GHz thermal emission corresponds to 20 mJy.  The faintest detectable
\ion{H}{2} regions correspond to the ionization by an O5 star if a
single ionizing source is assumed. We note that the detection limit of
\ion{H}{2} regions is quite good, $L(\mathrm{H}\alpha) = 2 \times 10^{36}$ ergs
s$^{-1}$, corresponding to one-fourth the luminosity of the Orion Nebula.

Using the first NANTEN CO survey ({\it Fukui et al.}, 1999; {\it
Mizuno et al.}, 2001b), the GMCs in the LMC were classified into the
three categories according to their associated young objects ({\it
Fukui et al.}, 1999; {\it Yamaguchi et al.}, 2001c):

%\begin{enumerate}
\newcounter{listcnt}
\begin{list}{\Roman{listcnt}}{\usecounter{listcnt}}
\item Starless GMCs (no early O stars); ``starless'' here indicates no
associated early O star capable of ionizing an \ion{H}{2} regions, which
does not exclude the possibility of associated young stars later than
B-type

\item GMCs with \ion{H}{2} regions only; those with small \ion{H}{2}
regions whose H$\alpha$ luminosity is less than $10^{37}$ erg s$^{-1}$

\item GMCs with \ion{H}{2} regions and stellar clusters; those with
stellar clusters and large \ion{H}{2} regions of H$\alpha$ luminosity
greater than $10^{37}$ erg s$^{-1}$.  \end{list}
%\end{enumerate}

The new NANTEN GMC catalog ({\it Fukui et al.}, 2006) is
used to improve and confirm the statistics of these three classes
({\it Kawamura et al.},~2006).  For the updated sample of 181 GMCs in
Fig.~\ref{lmcmap}, Fig.~\ref{histo} shows the frequency distribution of
the apparent separation of young objects, i.e., optical \ion{H}{2}
regions and stellar clusters, measured from the lowest contour of the
nearest GMC.  Obviously, the youngest stellar clusters, SWB0 and
\ion{H}{2} regions, exhibit marked peaks within 50-100 pc, indicating
their strong concentrations towards GMCs. Comparisons of these
distributions with a purely random distribution is shown by lines.
The differences between these peaks are significant.  The correlation
with young clusters establishes the physical association of the young
objects with the GMCs. On the other hand, clusters older than SWB I
show almost no correlation with GMCs.

\begin{figure}[t!] \epsscale{0.75} \plotone{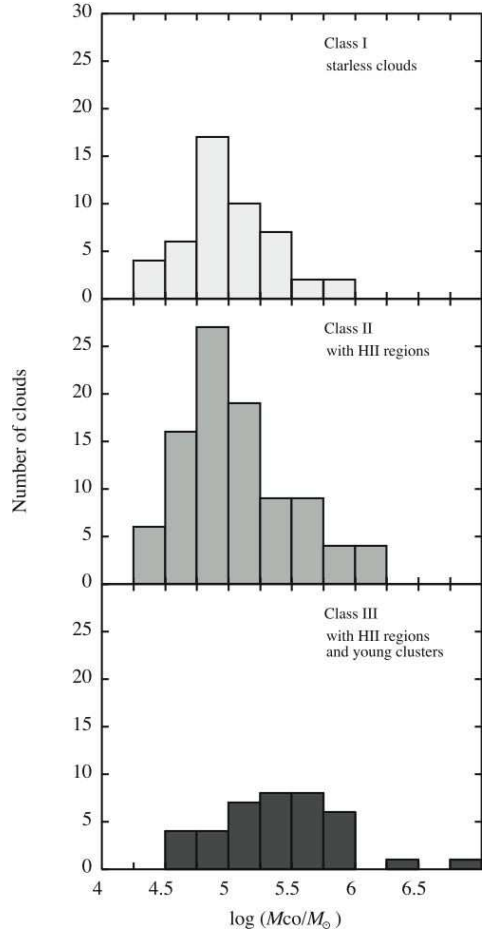}
\caption{\small  Histograms of the mass of class I ({\it Top}), class
II ({\it Middle}), and class III ({\it Bottom}), respectively. Mass is
derived by using \xco = 5.4 $\times 10^{20}$ cm$^{-2} (\Kkmpers)^{-1}$
(Table 2).  \label{mass_hist}}  \end{figure}

\begin{figure}[ht!] 
\epsscale{1} 
\plotone{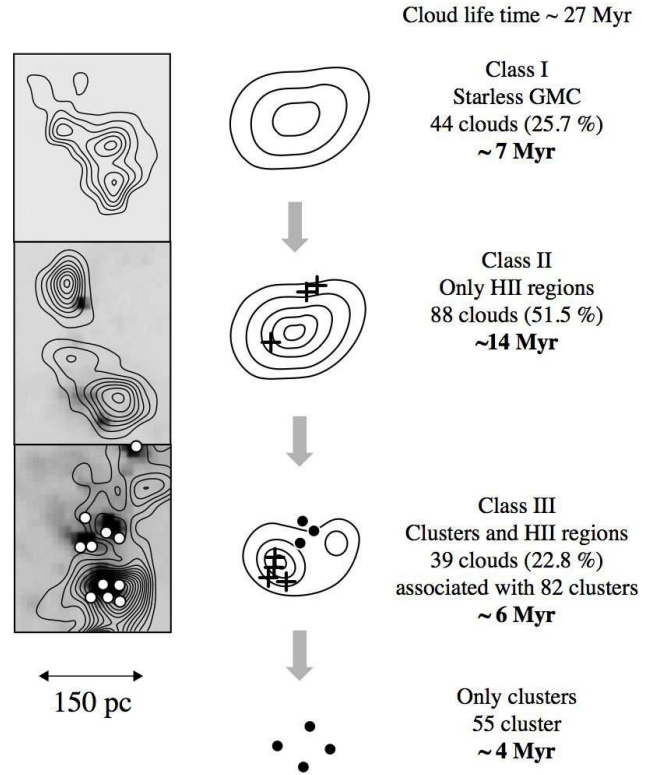}
\caption{\small  Evolutionary sequence of the GMCs in the LMC. An
example of the GMCs and illustration at each class are shown in the
left panels and the middle column, respectively. The images and
contours in the left panels are H$\alpha$ ({\it Kim et al.},~1999) and CO
integrated intensity by NANTEN ({\it Fukui et al.},~2001; 
{\it Fukui et al.},~in preparation);
contour levels are from \Kkmpers with 1.2 \Kkmpers intervals. Crosses
and filled circles indicate the position of the \ion{H}{2} regions and
young clusters, SWB 0 ({\it Bica et al.},~1996), respectively. The number of
the objects and the time scale at each class are also presented on the
right. \label{evolution}}  
\end{figure}

In order to look for any optically obscured \ion{H}{2} regions we have
also used the Parkes/ATNF radio continuum survey carried out at five
frequencies 1.4, 2.45, 4.75, 4.8, and 8.55 GHz ({\it Filipovic et
al.},~1995, 1998). The typical sensitivity limits of these new
datasets are quite good, allowing us to reach flux limits equivalent
to those in H${\alpha}$. The radio continuum results are summarized in
a catalog of 483 sources, and the spectral information makes it
possible to select
\ion{H}{2} regions and eliminate background sources not related the
LMC.  By comparing these data with the GMCs, we found that all of the
starless GMCs have no embedded \ion{H}{2} regions that are detectable
at radio wavelengths ({\it Kawamura et al.},~2006). 

Table \ref{timescale-table} summarizes the results of the present
comparison between GMCs and young objects, SWB0 clusters and the
\ion{H}{2} regions including radio sources. It shows that $\sim$ 25\%
of the GMCs are starless in the sense that they are not associated
with \ion{H}{2} regions or young clusters.  Fig.~\ref{mass_hist}
shows mass histograms of the three classes, I, II and III. These
indicate that the mass range of the three is from $10^{4.5}
M_{\odot}$ to a few times $10^{6} M_{\odot}$. It is also noteworthy
that class I GMCs tend to be less massive than the other two in the
sense that  the number of GMCs more massive than $10^{5} M_{\odot}$ is
about  half of those of class II and class III GMCs, respectively.

\begin{deluxetable}{lccl}
\tabletypesize{\small}
\setcounter{table}{3}
\tablecaption{Association of the young objects with GMCs
\label{timescale-table}}
\tablewidth{0pt}
\tablehead{\colhead{Class of GMC}  & \colhead{Number of GMCs\tablenotemark{a}} &
\colhead{Time scale\tablenotemark{b}} & \colhead{Association}\\
& & \colhead{(Myr)}}
\startdata
Class I & 44 (25.7 \%) & 7 & Starless \\
Class II & 88 (51.5 \%) & 14  & \ion{H}{2} regions \\
Class III  & 39 (22.8 \%) & 6 & \ion{H}{2} regions and clusters\tablenotemark{c}\\
\enddata
\tablenotetext{a}{GMCs with $M > 10^{4.5} M_{\sun}$; mass is derived by using
\xco = 5.4 $\times 10^{20}$ cm$^{-2} (\Kkmpers)^{-1}$ (Table 2).}
\tablenotetext{b}{A steady state evolution is assumed. The absolute time
scale is based on the age of stellar clusters; the age of SWB0 clusters, a
half of which are associated with the GMCs, is taken to be 10 Myr.}
\tablenotetext{c}{Young clusters or associations, SWB 0, by {\it Bica et al.}~(1996).}
\end{deluxetable} 

\bigskip
\noindent
\subsection{The Evolution of GMCs in the LMC}
\bigskip

The completeness of the present GMC sample covering the whole LMC
enables us to infer the evolutionary timescales of the GMCs.  We
assume a steady state evolution and therefore time spent in each phase
is proportional to the number of objects in Table
\ref{timescale-table}.  Fig.~\ref{evolution} is a scheme
representing the evolution suggested from Table \ref{timescale-table}.
The absolute time scale is based on the age of stellar clusters: the
age of SWB 0 clusters is taken to be 10 Myr. The first stage
corresponds to starless GMCs, having a long time scale of 7 Myr. This
is followed by a phase with small \ion{H}{2} regions, implying the
formation of a few to several O stars. The subsequent phase indicates
the most active formation of rich clusters including many early O
stars (one of such an example is N 159N). In the final phase, the GMC
has been more or less dissipated under the strong ionization and
stellar winds from O stars. The lifetime of a typical GMC in the LMC
is then estimated as the total of the timescales in Table
\ref{timescale-table}: $\sim$ 27 Myr, assuming that the GMC is
completely disrupted by the star formation. As noted earlier (Section
\ref{timescale}), the mass of class I GMCs tends to be smaller than
the rest. We may speculate that class I GMCs, and possibly part of
class II GMCs, are still growing in mass via mass accretion from their
surrounding lower density atomic gas.  In addition, the lifetime of
GMCs likely varies with cloud mass, so 27 Myr is only a characteristic
value and is probably uncertain by about 50\%.

\bigskip
\noindent
\subsection{\textbf{Star Formation in M 33}}
\bigskip

None of the other galaxies in our sample has as complete a record of
interstellar gas and star formation as does the LMC, which is due, in
part, to its proximity.  Nevertheless, it is possible to draw some
conclusions about the star formation in M33.  {\it Engargiola et
al.}~(2003) correlated the \ion{H}{2} regions cataloged by {\it Hodge
et al.}~(1999) with the 149 GMCs in the M33 catalog.  For reference,
the completeness limit of the {\it Hodge et al.}~(1999) cataloge is
$L(\mathrm{H}\alpha)=3\times 10^{35}$~erg~s$^{-1}$; a similar
range of \ion{H}{2} regions is cataloged in the LMC and M33.  {\it
Engargiola et al.}  (2003) assumed that an \ion{H}{2} region is
associated with a GMC if its boundary lies either within or tangent to
a GMC; 36\% of the flux from
\ion{H}{2} regions can be associated with the cataloged GMCs.
Correcting for the incompleteness of the GMCs cataloged below their
sensitivity limit suggests that $>90\%$ of the total flux of ionized
gas from M33 originates from GMCs.  Within the uncertainties,
essentially all of the flux from \ion{H}{2} regions is consistent with
an origin in GMCs.  Apparently, about half of the star formation in
M33 originates in GMCs below the the sensitivity limit of our survey.

A related question is to ask, what fraction of GMCs in M33 is actively
forming stars?  {\it Engargiola et al.}~(2003) counted the fraction of
GMCs with at least one \ion{H}{2} region having a separation $\Delta
r$.  They defined the correlation length, such that half the GMCs have
at least one \ion{H}{2}~region within this distance. The correlation
length for the GMCs and \ion{H}{2} regions is 35 pc; a random
distribution of GMCs and \ion{H}{2}~regions would return a correlation
length of 80 pc.  They assumed that a GMC is actively forming stars if
there is an \ion{H}{2}~region within 50 pc of the centroid of a GMC.
With this assumption, as many as 100 GMCs (67~\%) are forming massive
stars.  Of the 75 GMCs with masses above the median cataloged mass,
the fraction of clouds actively forming stars rises to 85\%.  They
estimate that the number of totally obscured \ion{H}{2}~regions affect
these results by at most 5\%.

Thus the fraction of GMCs without star formation is estimated to be
about 1/3, a fraction similar to that in the LMC.  The M33 study
estimated the lifetime of GMCs to be $\sim$ 20 Myr, also similar to
%estimated the lifetime of GMCs to be $\leq$ 20 Myr, also similar to
that found for the LMC.  The fraction of clouds without active star
formation is much higher than that found in the vicinity of the Sun
where only one of all of the GMCs within 2 kpc is found to be devoid
of star formation.  It is unclear whether this difference is
significant.  Neither the LMC analysis, nor the M33 analysis would
detect the low-mass star formation which it is proceeding in the
Taurus molecular clouds.  In any event, both the LMC and M33 studies
suggest that the fraction of clouds without star formation is small.
Thus the onset of star formation in GMCs is rather rapid not only in
the Milky Way, but in at least some lower mass spiral and irregular
galaxies.

\section{\textbf{FUTURE PROSPECTS}}
\bigskip

Studying GMCs in galaxies using CO emission requires spatial resolutions
higher than 30--40 pc. It will be possible to extend studies such as
ours to a few tens of Mpc soon, with the advent of ALMA and CARMA
arrays in the southern and northern hemispheres respectively. These
instruments will provide angular resolutions of 0.1--1 arcsec in
millimeter and sub-millimeter CO emission, corresponding 5--50 pc at 10 Mpc and will
provide unprecedented details of physical conditions in GMCs in galaxies. 
The work described in this chapter should be just the beginning of extragalactic GMC
studies.

\bigskip \section{\textbf {SUMMARY AND CONCLUSIONS}}

We have compared the properties of GMCs in 5 galaxies, four of which
have been surveyed in their entirety: the LMC, the SMC, M33, IC~10.
M31 was observed over a very limited area. The interstellar medium of 
all five galaxies is dominated by the atomic phase.

\begin{enumerate}

\item The GMCs do not, in general, show any relationship to the stellar
content of the galaxies except for the O stars
born in the GMCs.

\item There is a very good correlation between the locations of the
GMCs and filaments of \hi. Many filaments contain little of no
molecular gas even though they have similar surface densities compared
to those that are rich in GMCs. This suggests that clouds form from
the \ion{H}{1} rather than vice-versa.

\item There appears to be a clear evolutionary trend going from
filament formation $\rightarrow$ molecule formation $\rightarrow$ GMC
formation.  It is not clear however, whether the condensations that
form GMCs are first formed in the atomic filaments, or only after the
molecules have formed.

\item We derive \xco~for all of the galaxies assuming that the GMCs
are virialized.  Although there is some variation, a value of $\xco =
4 \times 10^{20}$ cm$^{-2}$ (K \kms)$^{-1}$ is a representative value
to within about 50\% except for the SMC, which has a value more than 3
times higher. There is no clear trend of
\xco~with metallicity.

\item The discrepancy between the virial value and the value
determined from $\gamma$-ray observations in the Milky Way suggests
that the GMCs are not virialized, if the $\gamma$-ray value is
applicable to other galaxies in the Local Group. In that case, a value
of $\xco = 2 \times 10^{20}$ cm$^{-2}$ (K \kms)$^{-1}$ may be more
appropriate.

\item The GMCs in our sample appear to satisfy the line width-size
relation for the Milky Way, but with an offset in the constant of
proportionality.  This offset may be due, at least in part, to the
different data analysis techniques for the MW and extragalactic data
sets.  For a given line width, the extragalactic clouds appear to be
about 50 \% larger.  Despite the systematic offset, there are small
but significant differences in the line width-size relationship among
GMCs in different galaxies.

\item The GMCs within a particular galaxy have a roughly constant
surface density.  If the value of \xco~we derive for each galaxy is
applied, the surface densities of the sample as a whole, have a
scatter of less than a factor of two.  

\item The mass spectra for the GMCs in all of the galaxies can be
characterized as a power law with a slope of $\sim -1.7$, with the
exception of M33, which has a slope of $-2.5$.

\item The ratio of \htwo~to \hi~on a pixel-by-pixel basis in galaxies
appears to be determined by the hydrostatic pressure in the disk.

\item About 1/4 -- 1/3 of the GMCs in the LMC and M33 appear to be
devoid of high-mass star formation.  

\item The association of stars and \hii~regions in the LMC suggests
a lifetime for the GMCs of about 27 Myr, with a quiescent phase that
is about 25\% of the age of the GMCs.  In M33, a lifetime of $\sim$ 20
Myr is measured.  For GMCs in these galaxies we estimate that typical
lifetimes are roughly 20--30 Myr. Both lifetimes are uncertain by
about 50\%.

\end{enumerate}

\bigskip 

\textbf{
Acknowledgments.}This work is partially supported by US National
Science Foundation under grants AST-0228963 and AST-0502605, a
Grant-in-Aid for Scientific Research from the Ministry of Education,
Culture, Sports, Science and Technology of Japan (No.\ 15071203), and
from JSPS (No.\ 14102003).  The NANTEN project is based on a mutual
agreement between Nagoya University and the Carnegie Institution of
Washington (CIW). We greatly appreciate the hospitality of all the
staff members of the Las Campanas Observatory of CIW. We are thankful
to many Japanese public donors and companies who contributed to the
realization of the project.  We would like to acknowledge
Drs. L. Stavely-Smith and M. Filipovic for the kind use of their
radio continuum data prior to publication.

\bigskip
%\newpage

\centerline\textbf{ REFERENCES}
\bigskip
\parskip=0pt
{\small
\baselineskip=11pt

\refs Allen R.~J.\ (2001) In {\em ASP Conf.~Ser.~240: Gas and Galaxy
Evolution 240}, (J. Hibbard and J. van Gorkom, eds.),
pp. 331-337. ASP Conf. Series, San Francisco.

\refs
Bica E., Claria J. J., Dottori H., Santos J. F. C. Jr., and Piatti A. E.,
(1996) \apjs, {\em 102}, 57-73.

\refs
Blitz L. and Rosolowsky E.\ (2004), \apj, {\em 612}, L29-L32.

\refs
Blitz L. and Rosolowsky E.\ (2006), submitted.

\refs
Bolatto A. et al. (2006), in preparation.

\refs
Braun R., Walterbos R.~A.~M., Kennicutt R.~C., and Tacconi L.~J.\
{1994}, \apj, {\em 420}, 558-569.

\refs
Cohen R.~S., Dame 
T.~M., Garay G., Montani J., Rubio M., and Thaddeus P.\ {1988}, \apj, 
{\em 331}, L95-L99.

\refs
Dame T., M., Hartmann D., and Thaddeus P.
(2001) \apj {\em 547}, 792-813.

\refs
Davies R. D., Elliott K. H., and Meaburn J.
(1976) {\em Mem. R. Astron. Soc.}, {\em 81}, 89-128.

\refs
de Boer K.~S., Braun J. M., Vallenari A., and Mebold U.~
(1998), \aap, {\em 329}, L49-L52.

\refs
Deul E. R. and van der Hulst J. M.
(1987) \aaps {\em 67}, 509-539.

\refs
Elmegreen B.~G.\ (1993), \apj, {\em 411}, 170-177.

\refs
Engargiola G., Plambeck R. L, Rosolowsky E., and Blitz L.
(2003) \apjs, {\em 149}, 343-363.

\refs
Filipovic M.~D., Haynes R.~F., White G.~L., Jones P.~A., Klein
U., and Wielebinski R.\ (1995), \aaps, {\em 111}, 311-332.

\refs
Filipovic M.~D., Haynes R.~F., White G.~L., and Jones P.~A.\ (1998),
\aaps, {\em 130}, 421-440.

\refs
Fujimoto M. and Noguchi M.\ (1990) {\em Publ. Astron. Soc. Japan,
42}, 505-516.

\refs
Fukui Y., Mizuno N., Yamaguchi R., Mizuno A., Onishi T., et al.~
(1999) {\em Publ. Astron. Soc. Japan, 51}, 745-749.

\refs
Fukui Y., Mizuno N., Yamaguchi R., Mizuno A., and Onishi T.
(2001) {\em Publ. Astron. Soc. Japan Letters, 53}, L41-L44.

\refs
Fukui Y. et al.\ (2006), in preparation.

\refs
Gao Y. and Solomon P.~M.\ (2004), \apj, {\em 606}, 271-290.

\refs
Gerola H. and Seiden P.~E. (1978) \apj, {\em 223} 129-135.

\refs 
Helfer T.~T. and Blitz L.\ (1993), \apj, {\em 419}, 86-93.

\refs
Helfer T.~T., Thornley M.~D., Regan M.~W., Wong T., Sheth K.,
Vogel S.~N., Blitz L., and Bock D.~C.-J.\ (2003), \apjs, {\em 145}, 259-327.

\refs
Henize K.~G.\ (1956), \apjs, {\em 2}, 315-344.

\refs
Henry R.~B.~C. and Howard J.~W.\ (1995), \apj, {\em 438}, 170-180.

\refs Heyer M.~H., Carpenter 
J.~M., and Snell R.~L.\ (2001), \apj, {\em 551}, 852-866.

\refs
Hodge P.~W., Balsley J., Wyder T.~K., and Skelton B.~P.\ (1999), \pasp,
{\em 111}, 685-690.

\refs 
Jarrett T.~H., 
Chester T., Cutri R., Schneider S.~E., and Huchra J.~P.\ (2003), \aj,
{\em 125}, 525-554.

\refs
Kawamura A. et al.\ (2006) in preparation.

\refs
Kennicutt R.~C.\ (1998), \apj, {\em 498}, 541-552.

\refs
Kennicutt R.\ C. Jr. and Hodge P.\ W.\
(1986) \apj, {\em 306}, 130-141.

\refs
Keto E., Ho L., and Lo K.Y. (2005), astro-ph/0508519.

\refs Kim S., Staveley-Smith 
L., Dopita M.~A., Freeman K.~C., Sault R.~J., Kesteven M.~J., and 
McConnell D. (1998), \apj, {\em 503}, 674-688.

\refs Kim S., Staveley-Smith 
L., Dopita M.~A., Sault R.~J., Freeman K.~C., Lee Y., and Chu Y.-H.
(2003), \apjs, {\em 148}, 473-486.

\refs
Kregel M., van der Kruit P.~C., and de Grijs R.\ {2002}, \mnras, {\em
334}, 646-668.

\refs
Koornneef J.\ (1984) In {\em IAU Symp. 108, Structure and Evolution of
the Magellanic Clouds}, (S. van den Bergh and K.\ S.\ de Boer, eds.),
pop. 333-339, D. Reidel, Dordrecht.

\refs
Lada C.~J., Margulis M., Sofue Y., Nakai N., and Handa T.\ (1988),
\apj, {\em 328}, 143-160.

\refs %IC10
Leroy A., Bolatto, A., Walter, F., and Blitz, L. (2006), \apj, in press.

\refs
Massey P., Hodge P.~W., Holmes S., Jacoby G., King N.~L., Olsen
K., Saha A., and Smith C.\ (2001), \baas, {\em 33}, 1496.

\refs
Mizuno N., Rubio M., 
Mizuno A., Yamaguchi R., Onishi T., and Fukui Y. (2001a), 
{\em Pub. of Astron. Soc. Japan, 53}, L45-L49.

\refs
Mizuno N., Yamaguchi R., Mizuno A., Rubio M., Abe R., Saito H., Onishi
T., Yonekura Y., Yamaguchi N., Ogawa H., and Fukui Y.
(2001b), {\em Pub. of Astron. Soc. Japan, 53}, 971-984.

\refs
Mizuno N. et al. (2006) in preparation.

\refs Narayanan D., Groppi C.~E., Kulesa C.~A., and Walker C.~K.\
(2005), \apj, {\em 630}, 269-279.

\refs Oka T., Hasegawa T., Sato F., Tsuboi M., Miyazaki A., and
Sugimoto M.\ (2001), \apj, {\em 562}, 348-362.

\refs Regan M. and Vogel S. (1994) \apj, {\it 434}, 536-545.

\refs
Rosolowsky E. (2006), \apj, submitted.

\refs
Rosolowsky E. and Leroy A. (2006), \pasp, in press.

\refs
Rosolowsky E. and Blitz L. (2005) \apj {\em 623} 826-845.

\refs
Rosolowsky E., Plambeck R., Engargiola G., and Blitz L.
(2003) \apj {\em 599} 258-274.

\refs
Rubio M., Garay G., Montani J., and Thaddeus P. (1991), \apj, 368, 173-177.

\refs
Solomon P. M., Rivolo A. R., Barrett J. and Yahil A.
(1987) \apj {\em 319}, 730-741.

\refs
Spergel D.~N. and Blitz L.\ (1992), \nat, {\em 357}, 665-667.

\refs
Stanimirovi\'c S., Staveley-Smith L., Dickey J.~M., Sault R.~J.,
and Snowden S.~L.\ (1999),
\mnras, {\em 302}, 417-436.

\refs
Stanimirovi\'c S., Staveley-Smith L., van der Hulst J.~M., Bontekoe T.~R., 
Kester D.~J.~M., and Jones P.~A.\ (2000), \mnras, {\em 315}, 791-807.

\refs 
Strong A.~W. and Mattox J.~R.\ (1996) \aap, {\em 308}, L21-L24.

\refs
van der Kruit P.~C. and Searle L.\ (1981a), \aap, {\em 95}, 105-115.

\refs
van der Kruit P.~C. and Searle L.\ (1981b), \aap, {\em 95}, 116-126.

\refs 
Vogel S.~N., Boulanger F., and Ball R.\ (1987), \apj, {\em 321}, L145-L149.

\refs
Wilcots E. M. and Miller B. W.
(1998) \aj {\em 116}, 2363-2394.

\refs
Williams J.~P. and
McKee C.~F.\ (1997), \apj, {\em 476}, 166-183.

\refs
Wilson C.~D.\ (1994), \apj, {\em 434}, L11-L14.

\refs 
Wilson C.~D. and Reid I.~N.\ (1991), \apj, {\em 366}, L11-L14.

\refs 
Wilson C.~D. and Rudolph A.~L.\ (1993), \apj, {\em 406}, 477-481.

\refs
Wilson C.\ D. and Scoville N.
(1990) \apj, {\em 363}, 435-450.

\refs
Wong T. and Blitz L.\ (2002), \apj, {\em 569}, 157-183.

\refs
Yamaguchi R., Mizuno N., Onishi T.,
Mizuno A., and Fukui Y.\ (2001a), 
\apj, {\em 553}, L185-L188.

\refs
Yamaguchi R., Mizuno N., Onishi T.,
Mizuno A., and Fukui Y.\ (2001b), 
{\em Pub. of Astron. Soc. Japan, 53}, 959-969.

\refs
Yamaguchi R., Mizuno N., Mizuno A.,
Rubio M., Abe R.,Saito H., Moriguchi
Y., Matsunaga L., Onishi T., Yonekura
Y., and Fukui Y.~
(2001c), {\em Pub. of Astron. Soc. Japan, 53}, 985-1001.

\end{document}